\documentclass{IOS-Book-Article}

\usepackage{amsfonts}
\normalfont
\usepackage{xcolor}
\usepackage{tikz}
\usepackage[draft]{hyperref}
\usepackage{latexsym}
\usepackage{isolatin1}
\usepackage{mymacros}
\usepackage{infrules}
\usepackage{smallverbatim}

\def\verbatimsize{\small}

\newenvironment{example}{\medskip\noindent{\bf Example.}}{\par\medskip}
\newtheorem{theorem}{Theorem}
\newtheorem{lemma}[theorem]{Lemma}

\def\den#1{[\![#1]\!]}
\def\becomes{\leftarrow}
\def\red{\rightarrow}
\def\redm{\stackrel{*}{\red}}
\def\redp{\stackrel{+}{\red}}
\def\eval#1#2#3{#2, #1 \Rightarrow #3}
\def\evalinf#1#2{#2, #1 \Rightarrow \infty}
\def\terminates{\Downarrow}
\def\diverges{\Uparrow {}}
\def\goeswrong{\Downarrow {\tt wrong}}
\def\res#1{\lfloor #1 \rfloor}
\def\bind#1#2#3{#1 \rhd (\lambda #2.~ #3)}
\def\interp{{\cal I}}
\def\comp{{\tt comp}} 

\def\behaviors{{\cal B}}
\def\triple#1#2#3{\{\,#1\,\}~#2~\{\,#3\,\}}
\def\Triple#1#2#3{[\,#1\,]~#2~[\,#3\,]}
\def\defequal{~~\stackrel{\hbox{\rm\scriptsize def}}{=}~~}
\def\agree#1#2#3{#1 \approx #2 : #3}
\def\union{\cup}
\def\aequal{\stackrel{.}{=}}
\def\alessthan{\stackrel{.}{<}}
\def\alesseq{\stackrel{.}{\le}}

\def\linkcolor{}
\hyperbaseurl{http://gallium.inria.fr/~xleroy/courses/Marktoberdorf-2009/}
\def\coq#1{\href{Sem.html\##1}{{\footnotesize\sf[#1]}}}

\begin{document}
\begin{frontmatter}

\title{Mechanized semantics}
\subtitle{with applications to program proof and compiler verification}
\runningtitle{Mechanized semantics}

\author{\fnms{Xavier} \snm{Leroy}}
\runningauthor{X. Leroy}
\address{INRIA Paris-Rocquencourt}

\begin{abstract}
The goal of this lecture is to show how modern theorem provers---in
this case, the Coq proof assistant---can be used to mechanize the specification
of programming languages and their semantics, and to reason over
individual programs and over generic program transformations,
as typically found in compilers.  The topics covered include:
operational semantics (small-step, big-step, definitional
interpreters); a simple form of denotational semantics; axiomatic
semantics and Hoare logic; generation of verification conditions, with
application to program proof; compilation to virtual machine code and
its proof of correctness; an example of an optimizing program
transformation (dead code elimination) and its proof of correctness.
\end{abstract}

\end{frontmatter}

\section*{Introduction}

The semantics of a programming language describe {\em mathematically}
the meaning of programs written in this language.  An example of use
of semantics is to define a programming language with much greater
precision than standard language specifications written in English.
(See for example the definition of Standard ML \cite{SML}.)  In turn,
semantics enable us to formally verify some programs, proving that
they satisfy their specifications.  Finally, semantics are also
necessary to establish the correctness of algorithms and
implementations that operate over programs: interpreters, compilers,
static analyzers (including type-checkers and bytecode verifiers),
program provers, refactoring tools, etc.

Semantics for nontrivial programming languages can be quite large and
complex, making traditional, on-paper proofs using these semantics
increasingly painful and unreliable.  Automatic theorem provers and
especially interactive proof assistants have great potential to
alleviate these problems and scale semantic-based techniques all the
way to realistic programming languages and tools.  Popular proof
assistants that have been successfully used in this area include
ACL2, Coq, HOL4, Isabelle/HOL, PVS and Twelf.

The purpose of this lecture is to introduce students to this booming
field of mechanized semantics and its applications to program proof
and formal verification of programming tools such as compilers.  Using
the prototypical IMP imperative language as a concrete example,
we will:
\begin{itemize}
\item mechanize various forms of operational and denotational
  semantics for this language and prove their equivalence (sections~1~and~2);
\item introduce axiomatic semantics (Hoare logic) and show how to
  provide machine assistance for proving IMP programs using a
  verification condition generator (section~3);
\item define a non-optimizing compiler from IMP to a virtual machine
  (a small subset of the Java virtual machine) and prove the
  correctness of this compiler via a semantic preservation argument
  (section~4);
\item illustrate optimizing compilation through the development and
  proof of correctness of a dead code elimination pass (section~5).
\end{itemize}
We finish with examples of recent achievements and ongoing
challenges in this area (section~6).

We use the Coq proof assistant to specify semantics and program
transformations, and conduct all proofs.  The best reference on
Coq is Bertot and Cast\'eran's book \cite{Bertot-Casteran-Coqart}, but
for the purposes of this lecture, Bertot's short tutorial
\cite{Bertot-hurry} is largely sufficient.  The Coq software and
documentation is available as free software at
\url{http://coq.inria.fr/}.
By lack of time, we will not attempt to teach how to conduct
interactive proofs in Coq (but see the two references above).
However, we hope that by the end of this lecture, students will be
familiar enough with Coq's specification language to be able to read
the Coq development underlying this lecture, and to write Coq
specifications for problems of their own interest.


\begin{figure}
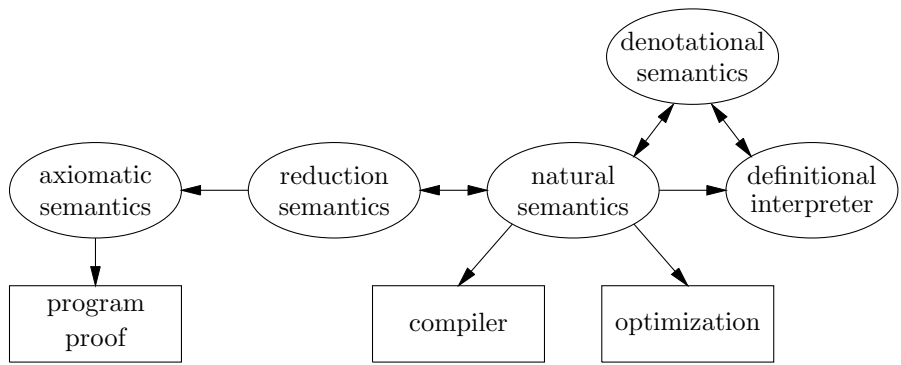


\begin{center}

\begin{pgfpicture}
  \pgfsetxvec{\pgfpoint{1.000in}{0in}}
  \pgfsetyvec{\pgfpoint{0in}{1.000in}}
  \begin{pgfscope}
    \pgfpathellipse{\pgfpointxy{0.450}{0.000}}{\pgfpointxy{0.450}{0}}{\pgfpointxy{0}{0.250}}
    \pgfusepath{stroke}
  \end{pgfscope}
  \pgftext[at=\pgfpointadd{\pgfpointxy{0.450}{0.000}}{\pgfpoint{0pt}{0.5 \baselineskip}}]{reduction}
  \pgftext[at=\pgfpointadd{\pgfpointxy{0.450}{0.000}}{\pgfpoint{0pt}{-0.5 \baselineskip}}]{semantics}
  \begin{pgfscope}
    \pgfpathellipse{\pgfpointxy{1.700}{0.000}}{\pgfpointxy{0.450}{0}}{\pgfpointxy{0}{0.250}}
    \pgfusepath{stroke}
  \end{pgfscope}
  \pgftext[at=\pgfpointadd{\pgfpointxy{1.700}{0.000}}{\pgfpoint{0pt}{0.5 \baselineskip}}]{natural}
  \pgftext[at=\pgfpointadd{\pgfpointxy{1.700}{0.000}}{\pgfpoint{0pt}{-0.5 \baselineskip}}]{semantics}
  \begin{pgfscope}
    \pgfpathellipse{\pgfpointxy{-0.800}{0.000}}{\pgfpointxy{0.450}{0}}{\pgfpointxy{0}{0.250}}
    \pgfusepath{stroke}
  \end{pgfscope}
  \pgftext[at=\pgfpointadd{\pgfpointxy{-0.800}{0.000}}{\pgfpoint{0pt}{0.5 \baselineskip}}]{axiomatic}
  \pgftext[at=\pgfpointadd{\pgfpointxy{-0.800}{0.000}}{\pgfpoint{0pt}{-0.5 \baselineskip}}]{semantics}
  \begin{pgfscope}
    \pgfpathellipse{\pgfpointxy{2.950}{0.000}}{\pgfpointxy{0.450}{0}}{\pgfpointxy{0}{0.250}}
    \pgfusepath{stroke}
  \end{pgfscope}
  \pgftext[at=\pgfpointadd{\pgfpointxy{2.950}{0.000}}{\pgfpoint{0pt}{0.5 \baselineskip}}]{definitional}
  \pgftext[at=\pgfpointadd{\pgfpointxy{2.950}{0.000}}{\pgfpoint{0pt}{-0.5 \baselineskip}}]{interpreter}
  \begin{pgfscope}
    \pgfpathellipse{\pgfpointxy{2.325}{0.700}}{\pgfpointxy{0.450}{0}}{\pgfpointxy{0}{0.250}}
    \pgfusepath{stroke}
  \end{pgfscope}
  \pgftext[at=\pgfpointadd{\pgfpointxy{2.325}{0.700}}{\pgfpoint{0pt}{0.5 \baselineskip}}]{denotational}
  \pgftext[at=\pgfpointadd{\pgfpointxy{2.325}{0.700}}{\pgfpoint{0pt}{-0.5 \baselineskip}}]{semantics}
  \begin{pgfscope}
    \pgfpathmoveto{\pgfpointxy{0.650}{-0.900}}
    \pgfpathlineto{\pgfpointxy{1.550}{-0.900}}
    \pgfpathlineto{\pgfpointxy{1.550}{-0.500}}
    \pgfpathlineto{\pgfpointxy{0.650}{-0.500}}
    \pgfpathlineto{\pgfpointxy{0.650}{-0.900}}
    \pgfusepath{stroke}
  \end{pgfscope}
  \pgftext[at=\pgfpointadd{\pgfpointxy{1.100}{-0.700}}{\pgfpoint{0pt}{-0.0 \baselineskip}}]{compiler}
  \begin{pgfscope}
    \pgfpathmoveto{\pgfpointxy{1.850}{-0.900}}
    \pgfpathlineto{\pgfpointxy{2.750}{-0.900}}
    \pgfpathlineto{\pgfpointxy{2.750}{-0.500}}
    \pgfpathlineto{\pgfpointxy{1.850}{-0.500}}
    \pgfpathlineto{\pgfpointxy{1.850}{-0.900}}
    \pgfusepath{stroke}
  \end{pgfscope}
  \pgftext[at=\pgfpointadd{\pgfpointxy{2.300}{-0.700}}{\pgfpoint{0pt}{-0.0 \baselineskip}}]{optimization}
  \begin{pgfscope}
    \pgfpathmoveto{\pgfpointxy{-1.250}{-0.900}}
    \pgfpathlineto{\pgfpointxy{-0.350}{-0.900}}
    \pgfpathlineto{\pgfpointxy{-0.350}{-0.500}}
    \pgfpathlineto{\pgfpointxy{-1.250}{-0.500}}
    \pgfpathlineto{\pgfpointxy{-1.250}{-0.900}}
    \pgfusepath{stroke}
  \end{pgfscope}
  \pgftext[at=\pgfpointadd{\pgfpointxy{-0.800}{-0.700}}{\pgfpoint{0pt}{0.5 \baselineskip}}]{program}
  \pgftext[at=\pgfpointadd{\pgfpointxy{-0.800}{-0.700}}{\pgfpoint{0pt}{-0.5 \baselineskip}}]{proof}
  \begin{pgfscope}
    \pgfpathmoveto{\pgfpointxy{1.000}{-0.025}}
    \pgfpathlineto{\pgfpointxy{0.900}{0.000}}
    \pgfpathlineto{\pgfpointxy{1.000}{0.025}}
    \pgfpathlineto{\pgfpointxy{1.000}{-0.025}}
    \pgfsetfillcolor{black}
    \pgfsetlinewidth{0.100pt}
    \pgfusepath{fill,stroke}
  \end{pgfscope}
  \begin{pgfscope}
    \pgfpathmoveto{\pgfpointxy{1.150}{0.025}}
    \pgfpathlineto{\pgfpointxy{1.250}{0.000}}
    \pgfpathlineto{\pgfpointxy{1.150}{-0.025}}
    \pgfpathlineto{\pgfpointxy{1.150}{0.025}}
    \pgfsetfillcolor{black}
    \pgfsetlinewidth{0.100pt}
    \pgfusepath{fill,stroke}
  \end{pgfscope}
  \begin{pgfscope}
    \pgfpathmoveto{\pgfpointxy{1.000}{0.000}}
    \pgfpathlineto{\pgfpointxy{1.150}{0.000}}
    \pgfusepath{stroke}
  \end{pgfscope}
  \begin{pgfscope}
    \pgfpathmoveto{\pgfpointxy{-0.250}{-0.025}}
    \pgfpathlineto{\pgfpointxy{-0.350}{0.000}}
    \pgfpathlineto{\pgfpointxy{-0.250}{0.025}}
    \pgfpathlineto{\pgfpointxy{-0.250}{-0.025}}
    \pgfsetfillcolor{black}
    \pgfsetlinewidth{0.100pt}
    \pgfusepath{fill,stroke}
  \end{pgfscope}
  \begin{pgfscope}
    \pgfpathmoveto{\pgfpointxy{0.000}{0.000}}
    \pgfpathlineto{\pgfpointxy{-0.250}{0.000}}
    \pgfusepath{stroke}
  \end{pgfscope}
  \begin{pgfscope}
    \pgfpathmoveto{\pgfpointxy{2.400}{0.025}}
    \pgfpathlineto{\pgfpointxy{2.500}{0.000}}
    \pgfpathlineto{\pgfpointxy{2.400}{-0.025}}
    \pgfpathlineto{\pgfpointxy{2.400}{0.025}}
    \pgfsetfillcolor{black}
    \pgfsetlinewidth{0.100pt}
    \pgfusepath{fill,stroke}
  \end{pgfscope}
  \begin{pgfscope}
    \pgfpathmoveto{\pgfpointxy{2.150}{0.000}}
    \pgfpathlineto{\pgfpointxy{2.400}{0.000}}
    \pgfusepath{stroke}
  \end{pgfscope}
  \begin{pgfscope}
    \pgfpathmoveto{\pgfpointxy{2.145}{0.385}}
    \pgfpathlineto{\pgfpointxy{2.225}{0.450}}
    \pgfpathlineto{\pgfpointxy{2.185}{0.355}}
    \pgfpathlineto{\pgfpointxy{2.145}{0.385}}
    \pgfsetfillcolor{black}
    \pgfsetlinewidth{0.100pt}
    \pgfusepath{fill,stroke}
  \end{pgfscope}
  \begin{pgfscope}
    \pgfpathmoveto{\pgfpointxy{2.098}{0.241}}
    \pgfpathlineto{\pgfpointxy{2.018}{0.177}}
    \pgfpathlineto{\pgfpointxy{2.059}{0.272}}
    \pgfpathlineto{\pgfpointxy{2.098}{0.241}}
    \pgfsetfillcolor{black}
    \pgfsetlinewidth{0.100pt}
    \pgfusepath{fill,stroke}
  \end{pgfscope}
  \begin{pgfscope}
    \pgfpathmoveto{\pgfpointxy{2.165}{0.370}}
    \pgfpathlineto{\pgfpointxy{2.079}{0.257}}
    \pgfusepath{stroke}
  \end{pgfscope}
  \begin{pgfscope}
    \pgfpathmoveto{\pgfpointxy{2.465}{0.355}}
    \pgfpathlineto{\pgfpointxy{2.425}{0.450}}
    \pgfpathlineto{\pgfpointxy{2.505}{0.385}}
    \pgfpathlineto{\pgfpointxy{2.465}{0.355}}
    \pgfsetfillcolor{black}
    \pgfsetlinewidth{0.100pt}
    \pgfusepath{fill,stroke}
  \end{pgfscope}
  \begin{pgfscope}
    \pgfpathmoveto{\pgfpointxy{2.591}{0.272}}
    \pgfpathlineto{\pgfpointxy{2.632}{0.177}}
    \pgfpathlineto{\pgfpointxy{2.552}{0.241}}
    \pgfpathlineto{\pgfpointxy{2.591}{0.272}}
    \pgfsetfillcolor{black}
    \pgfsetlinewidth{0.100pt}
    \pgfusepath{fill,stroke}
  \end{pgfscope}
  \begin{pgfscope}
    \pgfpathmoveto{\pgfpointxy{2.485}{0.370}}
    \pgfpathlineto{\pgfpointxy{2.571}{0.257}}
    \pgfusepath{stroke}
  \end{pgfscope}
  \begin{pgfscope}
    \pgfpathmoveto{\pgfpointxy{1.185}{-0.441}}
    \pgfpathlineto{\pgfpointxy{1.100}{-0.500}}
    \pgfpathlineto{\pgfpointxy{1.147}{-0.408}}
    \pgfpathlineto{\pgfpointxy{1.185}{-0.441}}
    \pgfsetfillcolor{black}
    \pgfsetlinewidth{0.100pt}
    \pgfusepath{fill,stroke}
  \end{pgfscope}
  \begin{pgfscope}
    \pgfpathmoveto{\pgfpointxy{1.382}{-0.177}}
    \pgfpathlineto{\pgfpointxy{1.166}{-0.425}}
    \pgfusepath{stroke}
  \end{pgfscope}
  \begin{pgfscope}
    \pgfpathmoveto{\pgfpointxy{2.253}{-0.408}}
    \pgfpathlineto{\pgfpointxy{2.300}{-0.500}}
    \pgfpathlineto{\pgfpointxy{2.215}{-0.441}}
    \pgfpathlineto{\pgfpointxy{2.253}{-0.408}}
    \pgfsetfillcolor{black}
    \pgfsetlinewidth{0.100pt}
    \pgfusepath{fill,stroke}
  \end{pgfscope}
  \begin{pgfscope}
    \pgfpathmoveto{\pgfpointxy{2.018}{-0.177}}
    \pgfpathlineto{\pgfpointxy{2.234}{-0.425}}
    \pgfusepath{stroke}
  \end{pgfscope}
  \begin{pgfscope}
    \pgfpathmoveto{\pgfpointxy{-0.775}{-0.400}}
    \pgfpathlineto{\pgfpointxy{-0.800}{-0.500}}
    \pgfpathlineto{\pgfpointxy{-0.825}{-0.400}}
    \pgfpathlineto{\pgfpointxy{-0.775}{-0.400}}
    \pgfsetfillcolor{black}
    \pgfsetlinewidth{0.100pt}
    \pgfusepath{fill,stroke}
  \end{pgfscope}
  \begin{pgfscope}
    \pgfpathmoveto{\pgfpointxy{-0.800}{-0.250}}
    \pgfpathlineto{\pgfpointxy{-0.800}{-0.400}}
    \pgfusepath{stroke}
  \end{pgfscope}
\end{pgfpicture}
\ifx\Setlineno\undefined\else\Setlineno=27\fi
\end{center}

\caption{The various styles of semantics considered in this lecture
  and their uses.  A double arrow denotes a semantic equivalence
  result.  A single arrow from $A$ to $B$ means that semantics $A$ is
  used to justify the correctness of $B$.}

\end{figure}

The reference material for this lecture is the Coq development
available at
\url{http://gallium.inria.fr/~xleroy/courses/Marktoberdorf-2009/}.  
These notes explain and recapitulate the definitions and main results
using ordinary mathematical syntax, and provides bibliographical
references.  To help readers make the connection with the Coq
development, the Coq names for the definitions and theorems are given
as bracketed notes, {\footnotesize\sf\linkcolor[like this]}.  In the PDF
version of the present document, available at the Web site above,
these notes are hyperlinks pointing directly to the corresponding Coq
definitions and theorems in the development.

\section{Symbolic expressions} \label{s:expressions}

\subsection{Syntax}

As a warm-up exercise, we start by formalizing the syntax and
semantics of a simple language of expressions comprising variables
($x$, $y$, \ldots), integer constants $n$, and two
operators $+$ and $-$.

\begin{syntax}
\syntaxclass{Expressions: \coq{expr}}
e & ::= & x \alt n \alt e_1 + e_2 \alt e_1 - e_2
\end{syntax}

The Coq representation of expressions is as a {\em inductive type},
similar to an ML or Haskell datatype.

\begin{verbatim}
Definition ident := nat.
Inductive expr : Type :=
  | Evar: ident -> expr
  | Econst: Z -> expr  
  | Eadd: expr -> expr -> expr 
  | Esub: expr -> expr -> expr.
\end{verbatim}

{\tt nat} and {\tt Z} are predefined types for natural numbers and integers,
respectively.  Each case of the inductive type is a function that
constructs terms of type {\tt expr}.  For instance, {\tt Evar} applied to the
name of a variable produces the representation of the corresponding
expression; and {\tt Eadd} applied to the representations of two
subexpressions $e_1$ and $e_2$ returns the representation of the
expression $e_1 + e_2$.  Moreover, all terms of type {\tt expr} are
finitely generated by repeated applications of the~4~constructor
functions; this enables definitions by pattern matching and reasoning
by case analysis and induction.

\subsection{Denotational semantics}

The simplest and perhaps most natural way to specify the semantics of
this language is as a function $\den{e}~s$ that associates an integer
value to the expression $e$ in the state $s$.  States associate values
to variables.  
$$\begin{array}{c}
\den{x}~s = s(x) \qquad
\den{n}~s = n \\[2mm]
\den{e_1 + e_2}~s = \den{e_1}~s + \den{e_2}~s \qquad
\den{e_1 - e_2}~s = \den{e_1}~s - \den{e_2}~s
\end{array}$$

In Coq, this denotational semantics is presented as a recursive
function \coq{eval\_expr}.
\begin{verbatim}
Definition state := ident -> Z.
Fixpoint eval_expr (s: state) (e: expr) {struct e} : Z :=
  match e with
  | Evar x => s x
  | Econst n => n
  | Eadd e1 e2 => eval_expr s e1 + eval_expr s e2
  | Esub e1 e2 => eval_expr s e1 - eval_expr s e2
  end.
\end{verbatim}
{\tt Fixpoint} marks a recursive function definition.  The {\tt struct\ e}
annotation states that it is structurally recursive on its {\tt e}
parameter, and therefore guaranteed to terminate.  The {\tt match...with}
construct represents pattern-matching on the shape of the expression
{\tt e}.

\subsection{Using the denotational semantics}

The {\tt eval{\char95}expr} function can be used as an interpreter, to evaluate
expressions in a known environment.  For example:
\begin{verbatim}
Eval compute in (
  let x : ident := O in
  let s : state := fun y => if eq_ident y x then 12 else 0 in
  eval_expr s (Eadd (Evar x) (Econst 1))).
\end{verbatim}
Coq prints ``{\tt 13\ :\ Z}''.  For additional performance, efficient
executable Caml code can also be generated automatically from the Coq
definition of {\tt eval{\char95}expr} using the extraction mechanism of Coq.

Another use of {\tt eval{\char95}expr} is to reason symbolically over expressions
in arbitrary states.  Consider the following claim:
\begin{verbatim}
Remark expr_add_pos:
  forall s x,
  s x >= 0 -> eval_expr s (Eadd (Evar x) (Econst 1)) > 0.
\end{verbatim}
Using the {\tt simpl} tactic of Coq, the goal reduces to a purely
arithmetic statement:
\begin{verbatim}
  forall s x, s x >= 0 -> s x + 1 > 0.
\end{verbatim}
which can be proved by standard arithmetic (the {\tt omega} tactic).

Finally, the denotation function {\tt eval{\char95}expr} can also be used to prove
``meta'' properties of the semantics.  For example, we can easily show
that the denotation of an expression is insensitive to values
of variables not mentioned in the expression.
\begin{verbatim}
Lemma eval_expr_domain:
  forall s1 s2 e,
  (forall x, is_free x e -> s1 x = s2 x) ->
  eval_expr s1 e = eval_expr s2 e.
\end{verbatim}
The proof is a simple induction on the structure of {\tt e}.  The
predicate {\tt is{\char95}free}, stating whether a variable occurs in an
expression, is itself defined as a recursive function:
\begin{verbatim}
Fixpoint is_free (x: ident) (e: expr) {struct e} : Prop :=
  match e with
  | Evar y => x = y
  | Econst n => False
  | Eadd e1 e2 => is_free x e1 \/ is_free x e2
  | Esub e1 e2 => is_free x e1 \/ is_free x e2
  end.
\end{verbatim}
As the {\tt Prop} annotation indicates, the result of this function is not
a data type but a logical formula.

\subsection{Variants}

The denotational semantics we gave above interprets the {\tt +} and {\tt -}
operators as arithmetic over mathematical integer.  We can easily
interpret them differently, for instance as signed, modulo $2^{32}$
arithmetic (as in Java):
\begin{verbatim}
Fixpoint eval_expr (s: state) (e: expr) {struct e} : Z :=
  match e with
  | Evar x => s x
  | Econst n => normalize n
  | Eadd e1 e2 => normalize (eval_expr s e1 + eval_expr s e2)
  | Esub e1 e2 => normalize (eval_expr s e1 - eval_expr s e2)
  end.
\end{verbatim}
Here, ${\tt normalize}~n$ is $n$ reduced modulo $2^{32}$ to the interval
$[-2^{31}, 2^{31})$.

We can also account for undefined expressions.  In practical
programming languages, the value of an expression can be undefined for
several reasons: if it mentions a variable that was not previously
defined; in case of overflow during an arithmetic operation; in case
of an integer division by 0; etc.  A simple way to account for
undefinedness is to use the {\tt option} type, as defined in Coq's
standard library.  This is a two-constructor inductive type with
{\tt None} meaning ``undefined'' and ${\tt Some}~n$ meaning ``defined and
having value~$n$''.

\begin{verbatim}
Definition state := ident -> option Z.
Fixpoint eval_expr (s: state) (e: expr) {struct e} : option Z :=
  match e with
  | Evar x => s x
  | Econst n => Some n
  | Eadd e1 e2 =>
      match eval_expr s e1, eval_expr s e2 with
      | Some n1, Some n2 => Some (n1 + n2)
      | _, _ => None
      end
  | Esub e1 e2 =>
      match eval_expr s e1, eval_expr s e2 with
      | Some n1, Some n2 => Some (n1 - n2)
      | _, _ => None
      end
  end.
\end{verbatim}

\subsection{Summary}

The approach we followed in this section---denotational semantics
represented as a Coq recursive function---is natural and convenient,
but limited by a fundamental aspect of Coq: all functions must be
terminating, so that they are defined everywhere by construction.  The
termination guarantee can come either by the fact that they are
structurally recursive (recursive calls are only done on strict
sub-terms of the argument, as in the case of {\tt eval{\char95}expr}), or by
Noetherian recursion on a well-founded ordering.  Consequently, the
approach followed in this section cannot be used to give semantics to
languages featuring general loops or general recursion.  As we now
illustrate with the IMP language, we need to move away from functional
presentations of the semantics
(where a function computes a result given a state and a term)
and adopt relational presentations instead
(where a ternary predicate relates a state, a term, and a result).

\section{The IMP language and its semantics} \label{s:semantics}

\subsection{Syntax}

The IMP language is a very simple imperative language with
structured control.  Syntactically, it extends the language
of expressions from section~\ref{s:expressions} with boolean
expressions (conditions) and commands (statements):
\begin{syntax}
\syntaxclass{Expressions: \coq{expr}}
e & ::= & x \alt n \alt e_1 + e_2 \alt e_1 - e_2
\syntaxclass{Boolean expressions: \coq{bool\_expr}}
b & ::= & e_1 = e_2 \alt e_1 < e_2
\syntaxclass{Commands: \coq{cmd}}
c & ::= & {\tt skip} \alt x := e \alt c_1; c_2 \alt
          {\tt if\ }b{\tt \ then\ }c_1{\tt \ else\ }c_2 \alt
          {\tt while\ }b{\tt \ do\ }c{\tt \ done}
\end{syntax}%
The semantics of boolean expressions is given in the denotational
style of section~\ref{s:expressions}, as a
function from states to booleans \coq{eval\_bool\_expr}.
\begin{eqnarray*}
\den{e_1 = e_2}~s &=& \cases{{\tt true} & if $\den{e_1}~s = \den{e_2}~s$; \cr
                           {\tt false} & otherwise. \cr}
\\[1mm]
\den{e_1 < e_2}~s &=& \cases{{\tt true} & if $\den{e_1}~s < \den{e_2}~s$; \cr
                           {\tt false} & otherwise. \cr}
\end{eqnarray*}

\subsection{Reduction semantics} \label{s:smallstep}

A standard way to give semantics to languages such as IMP, where
programs may not terminate, is reduction semantics, popularized by
Plotkin under the name ``structural operational semantics''
\cite{Plotkin-SOS}, and also called ``small-step semantics''.  It
builds on a reduction relation $(c, s) \red (c', s')$, meaning: in
initial state $s$, the command $c$ performs one elementary step of
computation, resulting in modified state $s'$ and residual
computations $c'$. \coq{red}

\begin{pannel}
\srulenumber{red\_assign}
(x := e,~s) \red ({\tt skip},~s[x \becomes \den{e}~s])
\end

\irulenumber{red\_seq\_left}
(c_1, s) \red (c_1', s)
---------------------
((c_1; c_2), ~s) \red ((c_1'; c_2), ~s')
\end

\srulenumber{red\_seq\_skip}
(({\tt skip}; c), ~s) \red (c, s) 
\end

\irulenumber{red\_if\_true}
\den{b}~s = {\tt true}
----------------------
(({\tt if\ }b{\tt \ then\ }c_1{\tt \ else\ }c_2), s) \red (c_1, s)
\end

\irulenumber{red\_if\_false}
\den{b}~s = {\tt false}
----------------------
(({\tt if\ }b{\tt \ then\ }c_1{\tt \ else\ }c_2), s) \red (c_2, s)
\end

\irulenumber{red\_while\_true}
\den{b}~s = {\tt true}
---------------------
(({\tt while\ }b{\tt \ do\ }c{\tt \ done}), s) \red ((c; {\tt while\ }b{\tt \ do\ }c{\tt \ done}), s)
\end

\irulenumber{red\_while\_false}
\den{b}~s = {\tt false}
---------------------
(({\tt while\ }b{\tt \ do\ }c{\tt \ done}), s) \red ({\tt skip}, s)
\end
\end{pannel}

The Coq translation of such a definition by inference rules is called
an inductive predicate.  Such predicates build on the same inductive
definition mechanisms that we already use to represent abstract syntax
trees, but the resulting logical object is a proposition (sort {\tt Prop})
instead of a data type (sort {\tt Type}).

The general recipe for translating inference rules to an inductive
predicate is as follows.  First, write each axiom and rule as a proper
logical formula, using implications and universal quantification over
free variables.  For example, the rule {\tt red{\char95}seq{\char95}left} becomes
\begin{verbatim}
  forall c1 c2 s c1' s',
      red (c1, s) (c1', s') ->
      red (Cseq c1 c2, s) (Cseq c1' c2, s')
\end{verbatim}
Second, give a name to each rule.  (These names are called
``constructors'', by analogy with data type constructors.)  Last, wrap
these named rules in an inductive predicate definition like the
following.
\begin{verbatim}
Inductive red: (cmd * state) -> (cmd * state) -> Prop :=
  | red_assign: forall x e s,
      red (Cassign x e, s) (Cskip, update s x (eval_expr s e))
  | red_seq_left: forall c1 c2 s c1' s',
      red (c1, s) (c1', s') ->
      red (Cseq c1 c2, s) (Cseq c1' c2, s')
  | red_seq_skip: forall c s,
      red (Cseq Cskip c, s) (c, s)
  | red_if_true: forall s b c1 c2,
      eval_bool_expr s b = true ->
      red (Cifthenelse b c1 c2, s) (c1, s)
  | red_if_false: forall s b c1 c2,
      eval_bool_expr s b = false ->
      red (Cifthenelse b c1 c2, s) (c2, s)
  | red_while_true: forall s b c,
      eval_bool_expr s b = true ->
      red (Cwhile b c, s) (Cseq c (Cwhile b c), s)
  | red_while_false: forall b c s,
      eval_bool_expr s b = false ->
      red (Cwhile b c, s) (Cskip, s).
\end{verbatim}
Each constructor of the definition is a theorem that lets us conclude
${\tt red\ }(c,s)~(c',s')$ when the corresponding premises hold.  Moreover,
the proposition ${\tt red\ } (c,s)~(c',s')$ holds only if it was derived by
applying these theorems a finite number of times (smallest fixpoint).
This provides us with powerful reasoning principles: by case analysis
on the last rule used, and by induction on a derivation.  Consider for
example the determinism of the reduction relation:
\begin{verbatim}
Lemma red_deterministic:
  forall cs cs1, red cs cs1 -> forall cs2, red cs cs2 -> cs1 = cs2.
\end{verbatim}
It is easily proved by induction on a derivation of {\tt red\ cs\ cs1}
and a case analysis on the last rule used to conclude {\tt red\ cs\ cs2}.

\medskip

From the one-step reduction relation, we can define the the behavior
of a command $c$ in an initial state $s$ is obtained by forming
sequences of reductions starting at $c,s$:
\begin{itemize}
\item Termination with final state $s'$, written $(c,s) \terminates s'$:
  finite sequence of reductions to {\tt skip}. \coq{terminates}
$$(c,s) \redm ({\tt skip}, s') $$
\item Divergence, written $(c,s) \diverges$: infinite sequence of reductions.
\coq{diverges}
$$ \forall c', forall s', ~ (c,s) \redm (c', s') \Rightarrow 
   \exists c'',\exists s'', ~(c', s') \red (c'', s'') $$
\item Going wrong, written $(c,s) \goeswrong$: finite sequence of
  reductions to an irreducible state that is not {\tt skip}.
\coq{goes\_wrong}
$$(c,s) \red \cdots \red (c', s') \not\red {} \mbox{ with $c \not= {\tt skip}$}$$
\end{itemize}

\subsection{Natural semantics} \label{s:natural-semantics}

An alternative to structured operational semantics is Kahn's natural
semantics \cite{Kahn-nat-sem}, also called big-step semantics.  Instead of
describing terminating executions as sequences of reductions, natural
semantics aims at giving a direct axiomatization of executions using
inference rules.  

To build intuitions for natural semantics, consider a terminating
reduction sequence for the command $c; c'$.
$$ ((c;c'),~s \red ((c_1;c'),~s_1) \red \cdots \red
 (({\tt skip};c'),~s_2) \red  (c',~s_2) \red \cdots \red  ({\tt skip},~s_3)
$$
It contains a terminating reduction sequence for $c$, of the form
$ (c,s) \redm ({\tt skip}, s_2) $,
followed by another terminating sequence for~$(c', s_2)$.  

The idea of natural semantics is to write inference rules that follow
this structure and define a predicate $\eval s c {s'}$, meaning ``in
initial state $s$, the command $c$ terminates with final state $s'$ ''.
\coq{exec}

\begin{pannel}

\srulenumber{exec\_skip} \eval s {{\tt skip}} s \end

\srulenumber{exec\_assign} \eval s {x := e} {s[x \becomes \den{e}~s]} \end

\irulenumber{exec\_seq}
   \eval {s} {c_1} {s_1} & \eval {s_1} {c_2} {s_2}
   -----------------------------------------
   \eval {s} {c_1; c_2} {s_2}
\end

\irulenumber{exec\_if}
  \eval {s} {c_1} {s'} \mbox{ if $\den{b}~s = {\tt true}$} \\
  \eval {s} {c_2} {s'} \mbox{ if $\den{b}~s = {\tt false}$}
  ------------------------------------------
  \eval {s} {({\tt if\ }b{\tt \ then\ }c_1{\tt \ else\ }c_2)} {s'}
\end

\irulenumber{exec\_while\_stop}
  \den{b}~s = {\tt false}
  ------------------------------------
  \eval {s} {{\tt while\ }b{\tt \ do\ }c{\tt \ done}} {s}
\end

\irulenumber{exec\_while\_loop}
  \den{b}~s = {\tt true} &
  \eval {s} {c} {s_1} &
  \eval {s_1} {{\tt while\ }b{\tt \ do\ }c{\tt \ done}} {s_2}
  ------------------------------------
  \eval {s} {{\tt while\ }b{\tt \ do\ }c{\tt \ done}} {s_2}
\end

\end{pannel}

We now have two different semantics for the same language.  A
legitimate question to ask is whether they are equivalent: do both
semantics predict the same ``terminates / diverges / goes wrong''
behaviors for any given program?  Such an equivalence result
strengthens the confidence we have in both semantics.  Moreover, it
enables us to use whichever semantics is more convenient to prove a
property of interest.
We first show an implication from natural semantics to terminating
reduction sequences.

\begin{theorem} \coq{exec\_terminates}
If $\eval s c {s'}$, then $(c, s) \redm ({\tt skip}, s')$.
\end{theorem}

The proof is a straightforward induction on a derivation of $\eval {s} {c}
{s'}$ and case analysis on the last rule used.  Here is a
representative case: $c = c_1;c_2$.  By hypothesis, $\eval {s} {c_1;c_2} {s'}$.
By inversion, we know that $\eval {s} {c_1} {s_1}$ and $\eval {s_1}
{c_2} {s'}$ for some intermediate state $s_1$.  Applying the induction
hypothesis twice, we obtain
$(c_1, s) \redm ({\tt skip}, s_1)$ and $(c_2, s_1) \redm ({\tt skip}, s')$.
A context lemma (proved separately by induction) shows that
$((c_1; c_2), s) \redm (({\tt skip}; c_2), s_1)$.
To obtain the expected result, all we need to do is to assemble the
reduction sequences together, using the transitivity of $\redm$:
$$ 
((c_1; c_2), s) \redm (({\tt skip}; c_2), s_1) \red (c_2, s_1) \redm ({\tt skip}, s')
$$

The converse implication (from terminating reduction sequences to
natural semantics) is more difficult.  The idea is to consider mixed
executions that start with some reduction steps and finish in one big
step using the natural semantics:
$$ (c_1, s_1) \red \cdots \red (c_i, s_i) \Rightarrow s' $$
We first show that the last reduction step can always be ``absorbed''
by the final big step:
\begin{lemma} \coq{red\_preserves\_exec}
If $(c,s) \red (c', s')$ and $\eval {s'} {c'} {s''}$, then $\eval s c {s''}$.
\end{lemma}

Combining this lemma with an induction on the sequence of reduction,
we obtain the desired semantic implication:

\begin{theorem} \coq{terminates\_exec}
If $(c, s) \redm ({\tt skip}, s')$, then $\eval s c {s'}$.
\end{theorem}

\subsection{Natural semantics for divergence}

Kahn-style natural semantics correctly characterize programs that
terminate, either normally (as in section~\ref{s:natural-semantics})
or by going wrong (through the addition of so-called error rules).
For a long time it was believed that natural semantics is unable to account
for divergence.  As observed by Grall and Leroy \cite{Leroy-Grall-coindsem},
this is not true: diverging executions can also be described in the
style of natural semantics, provided a coinductive definition
(greatest fixpoint) is used.  Define the infinite execution relation
$\evalinf s c$ (from initial state $s$, the command $c$
diverges). \coq{execinf}

\begin{pannel}

\irulenumberdouble{execinf\_seq\_left}
   \evalinf s {c_1}
   -----------------------------------------
   \evalinf s {c_1; c_2}
\end

\irulenumberdouble{execinf\_seq\_right}
   \eval {s} {c_1} {s_1} & \evalinf {s_1} {c_2}
   -----------------------------------------
   \evalinf {s} {c_1; c_2}
\end

\irulenumberdouble{execinf\_if}
  \evalinf {s} {c_1} \mbox{ if $\den{b}~s = {\tt true}$} \\
  \evalinf {s} {c_2} \mbox{ if $\den{b}~s = {\tt false}$}
  ------------------------------------------
  \evalinf {s} {{\tt if\ }b{\tt \ then\ }c_1{\tt \ else\ }c_2}
\end

\irulenumberdouble{execinf\_while\_body}
  \den{b}~s = {\tt true} &
  \evalinf {s} {c}
  ------------------------------------
  \evalinf {s} {{\tt while\ }b{\tt \ do\ }c{\tt \ done}}
\end

\irulenumberdouble{execinf\_while\_loop}
  \den{b}~s = {\tt true} &
  \eval {s} {c} {s_1} &
  \evalinf {s_1} {{\tt while\ }b{\tt \ do\ }c{\tt \ done}}
  ------------------------------------
  \evalinf {s} {{\tt while\ }b{\tt \ do\ }c{\tt \ done}}
\end

\end{pannel}

As denoted by the double horizontal bars, these rules must be
interpreted {\em coinductively} as a greatest fixpoint
\cite[section 2]{Leroy-Grall-coindsem}.  Equivalently, the coinductive
interpretation corresponds to conclusions of {\em possibly infinite}
derivation trees, while the inductive interpretation corresponds to
{\em finite} derivation trees.  Coq provides built-in support for
coinductive definitions of data types and predicates.

As in section~\ref{s:natural-semantics} and perhaps even more so here,
we need to prove an equivalence between the $\evalinf s c$ predicate
and the existence of infinite reduction sequences.  One
implication follows from the decomposition lemma below:

\begin{lemma} \coq{execinf\_red\_step}
If $\evalinf s c$, there exists $c'$ and $s'$
such that $(c,s) \red (c',s')$ and $\evalinf {s'} {c'}$.
\end{lemma}

A simple argument by coinduction, detailed in
\cite{Leroy-Grall-coindsem}, then concludes the expected implication:

\begin{theorem} \coq{execinf\_diverges}
If $\evalinf s c$, then $(c, s) \diverges$.
\end{theorem}

The reverse implication uses two inversion lemmas:
\begin{itemize}
\item If $((c_1; c_2),~s) \diverges$, either
$(c_1,s) \diverges$ or there exists $s'$ such that
$(c_1,s) \redm ({\tt skip},s')$ and $(c_2, s') \diverges$.

\item If $({\tt while\ }b{\tt \ do\ }c{\tt \ done},~s) \diverges$, 
then $\den{b}~s = {\tt true}$ and
either
$(c,s) \diverges$ or there exists $s'$ such that
$(c,s) \redm ({\tt skip},s')$ and 
$({\tt while\ }b{\tt \ do\ }c{\tt \ done},~s') \diverges$
\end{itemize}

These lemmas follow from determinism of the $\red$ relation and the
seemingly obvious fact that any reduction sequence is either infinite
or stops, after finitely many reductions, on an irreducible
configuration:
$$ \forall c,s,~~(c,s) \diverges ~\vee~
                 \exists c', \exists s', ~(c,s) \redm (c', s') \wedge
                 (c', s') \not\red {} $$
The property above cannot be proved in Coq's
constructive logic: such a constructive proof would be, in essence, a
program that decides the halting problem.  However, we can add the
law of excluded middle ($\forall P,~P \vee \neg P$) to Coq as an axiom,
without breaking logical consistency.  The fact above can easily be
proved from the law of excluded middle.

\begin{theorem} \coq{diverges\_execinf}
If $(c,s) \diverges$, then $\evalinf s c$.
\end{theorem}

\subsection{Definitional interpreter}

As mentioned at the end of section~1, we cannot write a Coq function
with type ${\tt cmd} \rightarrow {\tt state} \rightarrow {\tt state}$
that would execute a command and return its final state whenever the
command terminates: this function would not be total.
We can, however, define a Coq function  $\interp(n, c, s)$
that executes $c$ in initial state $s$, taking as extra argument a
natural number $n$ used to bound the amount of computation performed.
This function returns either $\res {s'}$ (termination with state $s'$)
or $\bot$ (insufficient recursion depth). \coq{interp}
{\setlength\mathindent{0em}
\begin{eqnarray*}
\interp(0, c, s) & = & \bot \\
\interp(n+1, {\tt skip}, s) & = & \res s \\
\interp(n+1, x := e, s) & = & \res{s[x \becomes \den{e}~s]~} \\
\interp(n+1, (c_1;c_2), s) & = &
  \bind {\interp(n, c_1, s)} {s'} {\interp(n, c_2, s')} \\
\interp(n+1, ({\tt if\ }b{\tt \ then\ }c_1{\tt \ else\ }c_2), s) & = &
  \interp(n, c_1, s) \mbox{ if $\den{b}~s = {\tt true}$} \\
\interp(n+1, ({\tt if\ }b{\tt \ then\ }c_1{\tt \ else\ }c_2), s) & = &
  \interp(n, c_2, s) \mbox{ if $\den{b}~s = {\tt false}$} \\
\interp(n+1, ({\tt while\ }b{\tt \ do\ }c{\tt \ done}), s) & = & \res s
  \mbox{ if  $\den{b}~s = {\tt false}$} \\
\interp(n+1, ({\tt while\ }b{\tt \ do\ }c{\tt \ done}), s) & = &
  \bind {\interp(n, c, s)} {s'} {\interp(n, {\tt while\ }b{\tt \ do\ }c{\tt \ done}, s')}\\
&&  \mbox{ if  $\den{b}~s = {\tt true}$}
\end{eqnarray*}}
The ``bind'' operator $\rhd$, reminiscent of monads in functional
programming, is defined by $\bot \rhd f = \bot$ and $\res{s} \rhd f =
f(s)$.

A crucial property of this definitional interpreter is that it is
monotone with respect to the maximal recursion depth $n$.
Evaluation results are ordered by taking
$\bot \le \res s$ \coq{res\_le}.

\begin{lemma} \coq{interp\_mon} (Monotonicity of $\interp$.)
If $n \le m$, then $\interp(n,c,s) \le \interp(m,c,s)$.
\end{lemma}

Exploiting this property, we can show partial correctness results of
the definitional interpreter with respect to the natural semantics:

\begin{lemma} \coq{interp\_exec}
If $\interp(n, c, s) = \res{s'}$, then $\eval s c {s'}$.
\end{lemma}

\begin{lemma} \coq{exec\_interp}
If $\eval s c {s'}$, there exists an $n$ such that
$\interp(n, c, s) = \res{s'}$.
\end{lemma}

\begin{lemma} \coq{execinf\_interp}
If $\evalinf s c$, then $\interp(n,c,s) = \bot$ for all $n$.
\end{lemma}

\subsection{Denotational semantics}

A simple form of denotational semantics \cite{Mosses-denot-sem} can be
obtained by ``letting $n$ goes to infinity'' in the definitional
interpreter.

\begin{lemma} \coq{interp\_limit\_dep}
For every $c$, there exists a function $\den{c}$ from states
to evaluation results such that
$ \forall s,~ \exists m,~ \forall n \ge m,~ \interp(n,c,s) = \den{c}~s$.
\end{lemma}

Again, this result cannot be proved in Coq's constructive logic and
requires the axiom of excluded middle and an axiom of description.

This denotation function $\den{c}$ satisfies the equations of
denotational semantics:
{\setlength\mathindent{0em}
\begin{eqnarray*}
\den{{\tt skip}}~s & = & \res s \\
\den{x := e}~s & = & \res {s[x \becomes \den{e}~s]} \\
\den{c_1; c_2}~s & = & \bind {\den{c_1}~s} {s'} {\den{c_2}~s'} \\
\den{{\tt if\ }b{\tt \ then\ }c_1{\tt \ else\ }c_2}~s & = & \den{c_1}~s
   \mbox{ if $\den{b}~s = {\tt true}$} \\
\den{{\tt if\ }b{\tt \ then\ }c_1{\tt \ else\ }c_2}~s & = & \den{c_2}~s
   \mbox{ if $\den{b}~s = {\tt false}$} \\
\den{{\tt while\ }b{\tt \ do\ }c{\tt \ done}}~s & = & \res s
  \mbox{ if  $\den{b}~s = {\tt false}$} \\
\den{{\tt while\ }b{\tt \ do\ }c{\tt \ done}}~s & = &
  \bind {\den{c}~s} {s'} {\den{{\tt while\ }b{\tt \ do\ }c{\tt \ done}}~s'}
  \mbox{ if  $\den{b}~s = {\tt true}$}
\end{eqnarray*}}
Moreover, $\den{{\tt while\ }b{\tt \ do\ }c{\tt \ done}}$ is the smallest function
from states to results that satisfies the last two equations.

Using these properties of $\den{c}$, we can show full equivalence
between the denotational and natural semantics.

\begin{theorem} \coq{denot\_exec} \coq{exec\_denot}
$\eval s c {s'}$ if and only if $\den{c}~s = \res{s'}$.
\end{theorem}

\begin{theorem} \coq{denot\_execinf} \coq{execinf\_denot}
$\evalinf s c$ if and only if $\den{c}~s = \bot$.
\end{theorem}

\subsection{Further reading}

The material presented in this section is inspired by Nipkow
\cite{Nipkow-Winskel-right} (in Isabelle/HOL, for the IMP language)
and by Grall and Leroy \cite{Leroy-Grall-coindsem}
(in Coq, for the call-by-value $\lambda$-calculus).

We followed Plotkin's ``SOS'' presentation \cite{Plotkin-SOS} of
reduction semantics, characterized by structural inductive rules such
as \coq{red\_seq\_left}.  An alternate presentation, based on reduction
contexts, was introduced by Wright and Felleisen
\cite{Felleisen-Wright} and is very popular to reason about type
systems \cite{TAPL}.  

Definitions and proofs by coinduction can be formalized in two ways:
as greatest fixpoints in a set-theoretic presentation \cite{Aczel1977}
or as infinite derivation trees in proof theory
\cite[chap. 13]{Bertot-Casteran-Coqart}.  Grall and Leroy
\cite{Leroy-Grall-coindsem} connect the two approaches.

The definitional interpreter approach was identified by Reynolds in
1972.  See \cite{Reynolds-definitional-revisited} for a historical
perspective.

The presentation of denotational semantics we followed avoids the
complexity of Scott domains.  Mechanizations of domain theory with
applications to denotational
semantics include Agerholm \cite{Agerholm-domain-93} (in HOL),
Paulin \cite{Paulin-Kahn-networks} (in Coq)
and Benton \etal \cite{Benton-domain-09} (in Coq).

\section{Axiomatic semantics and program verification}

Operational semantics as in section~\ref{s:semantics} focuses on
describing actual executions of programs.  In contrast, axiomatic
semantics (also called Hoare logic) focuses on verifying logical
assertions between the values of programs at various program points.
It is the most popular approach to proving the correctness of
imperative programs.

\subsection{Weak Hoare triples and their rules} \label{s:weak-triples}

Following Hoare's seminal work \cite{Hoare-logic},
we consider logical formulas of the
form $\triple P c Q$, meaning ``if precondition $P$ holds, the command
$c$ does not go wrong, and if it terminates, the postcondition $Q$
holds''.  Here, $P$ and $Q$ are arbitrary predicates over states.
A formula $\triple P c Q$ is called a {\em weak Hoare triple} (by
opposition with strong Hoare triples discussed in
section~\ref{s:strong-triples}, which guarantee termination as well).
We first define some useful operations over predicates:
$$\begin{array}{rcl@{\qquad}rcl}
P[x \becomes e] & \defequal & \lambda s.~P(s[x \becomes \den{e}~s]) &
P \wedge Q & \defequal & \lambda s.~P(s) \wedge Q(s) \\
b{\tt \ true} & \defequal & \lambda s.~\den{b}~s = {\tt true} &
P \vee Q & \defequal & \lambda s.~P(s) \vee Q(s) \\
b{\tt \ false} & \defequal & \lambda s.~\den{b}~s = {\tt false} &
P \Longrightarrow Q & \defequal & \forall s,~P(s) \Rightarrow Q(s)
\end{array}$$

The axiomatic semantics, that is, the set of legal triples
$\triple P c Q$, is defined by the following inference rules: \coq{triple}

\begin{pannel}

\srulenumber{triple\_skip} \triple P {{\tt skip}} P \end

\srulenumber{triple\_assign} \triple {P[x \becomes e]} {x := e} {P} \end

\irulenumber{triple\_seq}
  \triple{P}{c_1}{Q} & \triple{Q}{c_2}{R}
  --------------------------
  \triple{P}{c_1;c_2}{R}
\end

\irulenumber{triple\_if}
  \triple{b {\tt \ true} \wedge P}{c_1}{Q} &
  \triple{b {\tt \ false} \wedge P}{c_2}{Q}
  ---------------------------------------------------
  \triple{P}{{\tt if\ }b{\tt \ then\ }c_1{\tt \ else\ }c_2}{Q}
\end

\irulenumber{triple\_while}
  \triple{b{\tt \ true} \wedge P}{c}{P}
  ---------------------------------------------------
  \triple{P}{{\tt while\ }b{\tt \ do\ }c{\tt \ done}}{b{\tt \ false} \wedge P}
\end

\irulenumber{triple\_consequence}
  P \Longrightarrow P' & \triple{P'}{c}{Q'} & Q' \Longrightarrow Q
  ----------------------------------------------------
  \triple{P}{c}{Q}
\end

\end{pannel}

\begin{example}
The triple $ \triple {a = bq + r} {r := r - b; q := q + 1} {a = bq + r} $
is derivable from rules {\tt triple{\char95}assign}, {\tt triple{\char95}seq} and
{\tt triple{\char95}consequence} because the following logical equivalences hold:
\begin{eqnarray*}
(a = bq+r)[q \becomes q+1] & \Longleftrightarrow &
a = b(q + 1) + r 
\\
(a = b(q + 1) + r)[r \becomes r-b] & \Longleftrightarrow &
a = b(q+1) + (r-b) = bq+r
\end{eqnarray*}
\end{example}

\subsection{Soundness of the axiomatic semantics}

Intuitively, a weak Hoare triple $\triple{P}{c}{Q}$ is valid if for
all initial states $s$ such that $P~s$ holds, either $(c,s)$ diverges
or it terminates in a state $s'$ such that $Q~s'$ holds.  We capture
the latter condition by the predicate $(c,s) {\tt \ finally\ }Q$, defined
coinductively as: \coq{finally}
\begin{pannel}

\irulenumberdouble{finally\_done}
Q(s)
----------------------
({\tt skip}, s) {\tt \ finally\ } Q
\end

\irulenumberdouble{finally\_step}
(c,s) \red (c', s') & (c', s'){\tt \ finally\ }Q
------------------------------------------
(c,s){\tt \ finally\ }Q
\end

\end{pannel}

In an inductive interpretation, rule {\tt finally{\char95}step} could only be
applied a finite number of steps, and therefore 
$(c,s) {\tt \ finally\ }Q$ would be equivalent to
$\exists s',~(c,s) \redm ({\tt skip},s') \wedge Q(s')$.  In the
coinductive interpretation, rule {\tt finally{\char95}step} can also be applied
infinitely many times, capturing diverging executions as well.

The semantic interpretation $\den{\triple{P}{c}{Q}}$ of a triple is,
then, the proposition
$$ \forall s,~P~s \Longrightarrow (c,s){\tt \ finally\ }Q \qquad
   \mbox{\coq{sem\_triple}}$$

We now proceed to show that if $\triple P c Q$ is derivable, the
proposition $\den{\triple{P}{c}{Q}}$ above holds.  We start by some
lemmas about the {\tt finally} predicate.

\begin{lemma} \coq{finally\_seq}
If $(c_1, s){\tt \ finally\ }Q$ and $\den{\triple{Q}{c_2}{R}}$, then \\
$((c_1; c_2), s){\tt \ finally\ }R$.
\end{lemma}

\begin{lemma} \coq{finally\_while}
If $\den{\triple{b{\tt \ true} \wedge P}{c}{P}}$ then \\
$\den{\triple{P}{{\tt while\ }b{\tt \ do\ }c{\tt \ done}}{b{\tt \ false} \wedge P}}$.
\end{lemma}

\begin{lemma} \coq{finally\_consequence}
If $(c, s){\tt \ finally\ }Q$ and $Q \Longrightarrow Q'$, then \\
$(c, s){\tt \ finally\ }Q'$.
\end{lemma}

We can then prove the expected soundness result by a straightforward
induction on a derivation of $\triple{P}{c}{Q}$:

\begin{theorem} \coq{triple\_correct}
If $\triple{P}{c}{Q}$ can be derived by the rules of axiomatic
semantics, then $\den{\triple{P}{c}{Q}}$ holds.
\end{theorem}

\subsection{Generation of verification conditions} \label{s:vcgen}

In this section, we enrich the syntax of IMP commands with
an annotation on {\tt while} loops (to give the loop invariant) and an
${\tt assert}(P)$ command to let the user provide assertions. \coq{acmd}
\begin{syntax}
\syntaxclass{Annotated commands:}
c & ::= & {\tt while\ }b{\tt \ do\ }\{P\}~c{\tt \ done} & loop with invariant \\
  &\alt & {\tt assert}(P) & explicit assertion \\
  &\alt & \ldots & other commands as in IMP
\end{syntax}
Annotated commands can be viewed as regular commands by erasing the
$\{P\}$ annotation on loops and turning ${\tt assert}(P)$ to {\tt skip}. \coq{erase}

The {\tt wp} function computes the weakest (liberal) precondition for $c$
given a postcondition $Q$. \coq{wp}
{\setlength\mathindent{0em}
\begin{eqnarray*}
{\tt wp}({\tt skip}, Q) & = & Q \\
{\tt wp}(x := e, Q) & = & Q[x \becomes e] \\
{\tt wp}((c_1; c_2), Q) & = & {\tt wp}(c_1, {\tt wp}(c_2, Q)) \\
{\tt wp}(({\tt if\ }b{\tt \ then\ }c_1{\tt \ else\ }c_2), Q) & = &
(b {\tt \ true} \wedge {\tt wp}(c_1, Q)) \vee (b {\tt \ false} \wedge {\tt wp}(c_2, Q)) \\
{\tt wp}(({\tt while\ }b{\tt \ do\ }\{P\}~c{\tt \ done}), Q) & = & P \\
{\tt wp}({\tt assert}(P), Q) & = & P
\end{eqnarray*}}
With the same arguments, the {\tt vcg} function (verification condition
generator) computes a conjunction of implications that must hold for
the triple $\triple{{\tt wp}(c,Q)}{c}{Q}$ to hold. \coq{vcg}
\begin{eqnarray*}
{\tt vcg}({\tt skip}, Q) & = & T \\
{\tt vcg}(x := e, Q) & = & T \\
{\tt vcg}((c_1; c_2), Q) & = & {\tt vcg}(c_1, {\tt wp}(c_2, Q)) \wedge {\tt vcg}(c_2, Q) \\
{\tt vcg}(({\tt if\ }b{\tt \ then\ }c_1{\tt \ else\ }c_2), Q) & = &
                            {\tt vcg}(c_1, Q) \wedge {\tt vcg}(c_2, Q) \\
{\tt vcg}(({\tt while\ }b{\tt \ do\ }\{P\}~c{\tt \ done}), Q) & = & {\tt vcg}(c, P) \\
& & {} \wedge (b {\tt \ false} \wedge P \Longrightarrow Q) \\
& & {} \wedge (b {\tt \ true} \wedge P \Longrightarrow {\tt wp}(c, P)) \\
{\tt vcg}({\tt assert}(P), Q) & = & P \Longrightarrow Q
\end{eqnarray*}

\begin{lemma} \coq{vcg\_correct}
If ${\tt vcg}(c, Q)$ holds, then $\triple{{\tt wp}(c, Q)}{c}{Q}$ can be
derived by the rules of axiomatic semantics.
\end{lemma}

The derivation of a Hoare triple $\triple{P}{c}{Q}$ can therefore be
reduced to the computation of the following ${\tt vcgen}(P,c,Q)$ logical
formula, and its proof. \coq{vcgen}

$$ {\tt vcgen}(P,c,Q) \defequal (P \Longrightarrow {\tt wp}(c,Q)) \wedge {\tt vcg}(c,Q) $$

\begin{theorem} \coq{vcgen\_correct} \label{vcgen-correct}
If ${\tt vcgen}(P, c, Q)$ holds, then $\triple{P}{c}{Q}$ can be
derived by the rules of axiomatic semantics.
\end{theorem}

\begin{example} Consider the following annotated IMP program $c$:
\begin{alltt}
    r := a; q := 0;
    while b < r+1 do \{\(I\)\} r := r - b; q := q + 1 done
\end{alltt}
and the following precondition $P$, loop invariant $I$ and
postcondition $Q$:
\begin{eqnarray*}
P & \defequal & \lambda s.~s({\tt a}) \ge 0 \wedge s({\tt b}) > 0 \\
I & \defequal & \lambda s.~s({\tt r}) \ge 0 \wedge s({\tt b}) > 0 \wedge 
          s({\tt a}) = s({\tt b}) \times s({\tt q}) + s({\tt r}) \\
Q & \defequal & \lambda s.~s({\tt q}) = s({\tt a}) / s({\tt b})
\end{eqnarray*}
To prove that $\triple {P}{c}{Q}$, we apply
theorem~\ref{vcgen-correct}, then ask Coq to compute and simplify the formula
${\tt vcgen}(P, c, Q)$.  We obtain the conjunction of three implications:
$$\begin{array}{l}
s({\tt a}) \ge 0 \wedge s({\tt b}) > 0
\Longrightarrow
s({\tt a}) \ge 0 \wedge s({\tt b}) > 0 \wedge s({\tt a}) = s({\tt b}) \times 0 + s({\tt a})
\\[2mm]
\neg(s({\tt b}) < s({\tt r})+1) \wedge s({\tt r}) \ge 0 \wedge s({\tt b}) > 0 \wedge 
s({\tt a}) = s({\tt b}) \times s({\tt q}) + s({\tt r}) \\
\qquad {} \Longrightarrow 
s({\tt q}) = s({\tt a}) / s({\tt b}) 
\\[2mm]
s({\tt b}) < s({\tt r})+1 \wedge s({\tt r}) \ge 0 \wedge s({\tt b}) > 0 \wedge 
s({\tt a}) = s({\tt b}) \times s({\tt q}) + s({\tt r}) \\
\qquad {}\Longrightarrow
s({\tt r}) - s({\tt b}) \ge 0 \wedge s({\tt b}) > 0 \wedge 
s({\tt a}) = s({\tt b}) \times (s({\tt q}) + 1) + (s({\tt r}) - s({\tt b}))
\end{array}$$
which are easy to prove by purely arithmetic reasoning.

\end{example}

\subsection{Strong Hoare triples} \label{s:strong-triples}

The axiomatic semantics we have seen so far enables us to prove
partial correctness properties of programs, but not their
termination.  To prove termination as well, we need to use strong
Hoare triples $\Triple P c Q$, meaning ``if precondition $P$ holds,
the command $c$ terminates and moreover the postcondition $Q$ holds''.

The rules defining valid strong Hoare triples are similar to those for
weak triples, with the exception of the {\tt while} rule, which contains
additional requirements that ensure termination of the loop. \coq{Triple}

\begin{pannel}

\srulenumber{Triple\_skip} \Triple P {{\tt skip}} P \end

\srulenumber{Triple\_assign} \Triple {P[x \becomes e]} {x := e} {P} \end

\irulenumber{Triple\_seq}
  \Triple{P}{c_1}{Q} & \Triple{Q}{c_2}{R}
  --------------------------
  \Triple{P}{c_1;c_2}{R}
\end

\irulenumber{Triple\_if}
  \Triple{b {\tt \ true} \wedge P}{c_1}{Q} &
  \Triple{b {\tt \ false} \wedge P}{c_2}{Q}
  ---------------------------------------------------
  \Triple{P}{{\tt if\ }b{\tt \ then\ }c_1{\tt \ else\ }c_2}{Q}
\end

\irulenumber{Triple\_while}
  (\forall v \in \mathbb{Z},~
         \Triple{b{\tt \ true} \wedge e_m \aequal v \wedge P}
                {c}
                {0 \alesseq e_m \alessthan v \wedge P}~)
  ---------------------------------------------------
  \Triple{P}{{\tt while\ }b{\tt \ do\ }c{\tt \ done}}{b{\tt \ false} \wedge P}
\end

\irulenumber{Triple\_consequence}
  P \Longrightarrow P' & \Triple{P'}{c}{Q'} & Q' \Longrightarrow Q
  ----------------------------------------------------
  \Triple{P}{c}{Q}
\end

\end{pannel}

In the {\tt Triple{\char95}while} rule, $e_m$ stands for an expression whose value
should decrease but remain nonnegative at each iteration.  The
precondition $e_m \aequal v$ and the postcondition
$0 \alesseq e_m \alessthan v$
capture this fact:
$$
e_m \aequal v \defequal \lambda s.~\den{e_m}~s = v
\qquad \qquad
0 \alesseq e_m \alessthan v \defequal \lambda s.~0 \le \den{e_m}~s < v
$$
The $v$ variable therefore denotes the value of the measure expression
at the beginning of the loop body.  Since it is not statically known
in general, rule {\tt Triple{\char95}while} quantifies universally over every
possible $v \in \mathbb{Z}$.  Conceptually, rule {\tt Triple{\char95}while} has
infinitely many premises, one for each possible value of $v$.  Such
infinitely branching inference rules cause no difficulty in Coq.

Note that the {\tt Triple{\char95}while} rule above is not powerful
enough to prove termination for some loops that occur in practice, for
example if the termination argument is based on a lexicographic
ordering.  A more general version of the rule could involve an
arbitrary well-founded ordering between states.

The semantic interpretation $\den{\Triple{P}{c}{Q}}$ of a strong Hoare
triple is the proposition
$$ \forall s,~P~s \Longrightarrow \exists s',~(\eval s c {s'}) \wedge Q(s')
   \qquad \mbox{\coq{sem\_Triple}}$$

As previously done for weak triples, we now prove the soundness of the
inference rules for strong triples with respect to this semantic
interpretation.

\begin{theorem} \coq{Triple\_correct}
If $\Triple{P}{c}{Q}$ can be derived by the rules of axiomatic
semantics, then $\den{\Triple{P}{c}{Q}}$ holds.
\end{theorem}

The proof is by an outer induction on a derivation of  $\Triple{P}{c}{Q}$
followed, in the {\tt while} case, by an inner induction on the value of
the associated measure expression.

\subsection{Further reading}

The material in this section follows Nipkow \cite{Nipkow-Winskel-right}
(in Isabelle/HOL) and Bertot \cite{Bertot-progsem} (in Coq),
themselves following Gordon \cite{Gordon-prog-logics-88}.

Separation logic \cite{OHearn-Reynolds-Yang-01,Reynolds-02} extends
axiomatic semantics with a notion of local reasoning: assertions carry
a {\em domain} (in our case, a set of variable; in pointer programs, a
set of store locations) and the logic enforces that nothing outside
the domain of the triple changes during execution.  Examples of
mechanized separation logics include
Marti \etal \cite{Marti-Affeldt-06} in Coq,
Tuch \etal \cite{Tuch-Klein-Norrish} in Isabelle/HOL,
Appel and Blazy \cite{Appel-Blazy-07} in Coq,
and Myreen and Gordon \cite{Myreen-07} in HOL4.

The generation of verification conditions (section~\ref{s:vcgen})
is an instance of a more general technique known as ``proof by
reflection'', which aims at replacing deduction steps by computations
\cite[chap.~16]{Bertot-Casteran-Coqart}.
The derivation of $\triple P c Q$ from the rules of
section~\ref{s:weak-triples} (a nonobvious process involving
nondeterminstic proof search) is replaced by the computation of
${\tt vcgen}(P,c,Q)$ (a trivial evaluation of a recursive function
application).  Proofs by reflection can tremendously speed up the
verification of combinatorial properties, as illustrated by Gonthier
and Werner's mechanized proof of the 4-color theorem \cite{Gonthier-4color}.

\section{Compilation to a virtual machine} \label{s:compilation}

There are several ways to execute programs:
\begin{itemize}
\item Interpretation: a program (the interpreter) traverses the
abstract syntax tree of the program to be executed, performing the
intended computations on the fly.
\item Compilation to native code: before execution, the program is
translated to a sequence of machine instructions.  These
instructions are those of a real microprocessor and are executed in
hardware.
\item Compilation to virtual machine code: before execution, the program is translated to a sequence of
instructions, These instructions are those of a {\em virtual
machine}.  They do not correspond to that of an existing hardware
processor, but are chosen close to the basic operations of the source
language.  Then, the virtual machine code is either interpreted (more
efficiently than source-level interpretation) or further translated to
real machine code.
\end{itemize}
In this section, we study the compilation of the IMP language to an
appropriate virtual machine.

\subsection{The IMP virtual machine}

A state of the machine is composed of: \coq{machine\_state}
\begin{itemize}
\item A fixed code $C$ (a list of instructions).
\item A variable program counter $pc$ (an integer position in $C$).
\item A variable stack $\sigma$ (a list of integers).
\item A store $s$ (mapping variables to integers).
\end{itemize}
The instruction set is as follows: \coq{instruction} \coq{code}
\begin{syntax}
i & ::=  & {\tt const}(n) & push $n$ on stack \\
  & \alt & {\tt var}(x) & push value of $x$ \\
  & \alt & {\tt setvar}(x) & pop value and assign it to $x$ \\
  & \alt & {\tt add} & pop two values, push their sum \\
  & \alt & {\tt sub} & pop two values, push their difference \\
  & \alt & {\tt branch}(\delta) & unconditional jump \\
  & \alt & {\tt bne}(\delta) & pop two values, jump if $\not =$ \\
  & \alt & {\tt bge}(\delta) & pop two values, jump if $\ge$ \\
  & \alt & {\tt halt} & end of program
\end{syntax}
In branch instructions, $\delta$ is an offset relative to the next
instruction.

The dynamic semantics of the machine is given by the following
one-step transition relation \coq{transition}.
$C(pc)$ is the instruction at position $pc$ in $C$, if any.
$$\begin{array}{ll}
C |- (pc, \sigma, s) \red (pc+1, n.\sigma, s)
& \mbox{if $C(pc) = {\tt const}(n)$} \\
C |- (pc, \sigma, s) \red (pc+1, s.(x).\sigma, s)
& \mbox{if $C(pc) = {\tt var}(n)$} \\
C |- (pc, n.\sigma, s) \red (pc+1, \sigma, s[x \becomes n])
& \mbox{if $C(pc) = {\tt setvar}(x)$} \\
C |- (pc, n_2.n_1.\sigma, s) \red (pc+1, (n_1+n_2).\sigma, s)
& \mbox{if $C(pc) = {\tt add}$} \\
C |- (pc, n_2.n_1.\sigma, s) \red (pc+1, (n_1-n_2).\sigma, s)
& \mbox{if $C(pc) = {\tt sub}$} \\
C |- (pc, \sigma, s) \red (pc+1+\delta, \sigma, s)
& \mbox{if $C(pc) = {\tt branch}(\delta)$} \\
C |- (pc, n_2.n_1.\sigma, s) \red (pc+1+\delta, \sigma, s)
& \mbox{if $C(pc) = {\tt bne}(\delta)$ and $n_1 \not= n_2$} \\
C |- (pc, n_2.n_1.\sigma, s) \red (pc+1, \sigma, s)
& \mbox{if $C(pc) = {\tt bne}(\delta)$ and $n_1 = n_2$} \\
C |- (pc, n_2.n_1.\sigma, s) \red (pc+1+\delta, \sigma, s)
& \mbox{if $C(pc) = {\tt bge}(\delta)$ and $n_1 \ge n_2$} \\
C |- (pc, n_2.n_1.\sigma, s) \red (pc+1, \sigma, s)
& \mbox{if $C(pc) = {\tt bge}(\delta)$ and $n_1 < n_2$}
\end{array}$$

As in section~\ref{s:smallstep}, the observable behavior of a machine
program is defined by sequences of transitions:
\begin{itemize}
\item Termination $C |- (pc, \sigma, s) \terminates s'$ if \\
$C |- (pc, \sigma, s) \redm (pc', \sigma', s')$ and $C(pc') = {\tt halt}$.
\item Divergence $C |- (pc, \sigma, s) \diverges$ if the machine makes
  infinitely many transitions from $(pc, \sigma, s)$.
\item Going wrong, otherwise.
\end{itemize}

\begin{example} The table below depicts the first 4 transitions of the
  execution of the code ${\tt var}(x); {\tt const}(1); {\tt add}; {\tt setvar}(x);
  {\tt branch}(-5)$.

{
\def\str{x \mapsto}
\normalbar
\def\instr#1{\multicolumn{1}{l}{#1}}

$$\begin{array}{l@{~}|@{~}l|@{~}l|@{~}l|@{~}l|@{~}l}
\mbox{stack} & \epsilon  & 12.\epsilon & 1.12.\epsilon
                                         & 13.\epsilon         & \epsilon
\\[1mm]
\mbox{store} & \str12    & \str12      & \str12 & \str12       & \str13
\\[1mm]
\mbox{p.c.}  & 0         & 1           & 2      & 3            & 4
\\[1mm]
\mbox{code}  & \instr{{\tt var}(x);}
                  & \instr{{\tt const}(1);}
                                & \instr{{\tt add};}
                                         & \instr{{\tt setvar}(x);}
                                                        & \instr{{\tt branch}(-5)}
\end{array}$$
}
The fifth transition executes the ${\tt branch}(-5)$ instruction, setting
the program counter back to 0.  The overall effect is that of an
infinite loop that increments $x$ by 1 at each iteration.
\end{example}

\subsection{The compilation scheme}

\begin{figure}
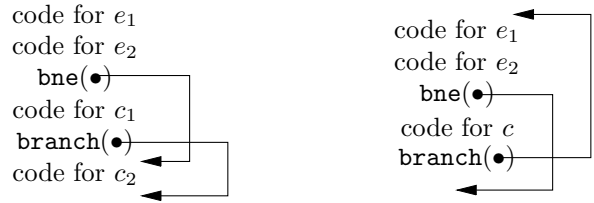


\begin{center}

\begin{pgfpicture}
  \pgfsetxvec{\pgfpoint{1.000in}{0in}}
  \pgfsetyvec{\pgfpoint{0in}{1.000in}}
  \pgftext[at=\pgfpointadd{\pgfpointxy{0.375}{0.000}}{\pgfpoint{0pt}{2.5 \baselineskip}}]{code for $e_1$}
  \pgftext[at=\pgfpointadd{\pgfpointxy{0.375}{0.000}}{\pgfpoint{0pt}{1.5 \baselineskip}}]{code for $e_2$}
  \pgftext[at=\pgfpointadd{\pgfpointxy{0.375}{0.000}}{\pgfpoint{0pt}{0.5 \baselineskip}}]{${\tt bne}(\bullet)$}
  \pgftext[at=\pgfpointadd{\pgfpointxy{0.375}{0.000}}{\pgfpoint{0pt}{-0.5 \baselineskip}}]{code for $c_1$}
  \pgftext[at=\pgfpointadd{\pgfpointxy{0.375}{0.000}}{\pgfpoint{0pt}{-1.5 \baselineskip}}]{${\tt branch}(\bullet)$}
  \pgftext[at=\pgfpointadd{\pgfpointxy{0.375}{0.000}}{\pgfpoint{0pt}{-2.5 \baselineskip}}]{code for $c_2$}
  \pgftext[at=\pgfpointadd{\pgfpointxy{2.375}{0.000}}{\pgfpoint{0pt}{2.0 \baselineskip}}]{code for $e_1$}
  \pgftext[at=\pgfpointadd{\pgfpointxy{2.375}{0.000}}{\pgfpoint{0pt}{1.0 \baselineskip}}]{code for $e_2$}
  \pgftext[at=\pgfpointadd{\pgfpointxy{2.375}{0.000}}{\pgfpoint{0pt}{-0.0 \baselineskip}}]{${\tt bne}(\bullet)$}
  \pgftext[at=\pgfpointadd{\pgfpointxy{2.375}{0.000}}{\pgfpoint{0pt}{-1.0 \baselineskip}}]{code for $c$}
  \pgftext[at=\pgfpointadd{\pgfpointxy{2.375}{0.000}}{\pgfpoint{0pt}{-2.0 \baselineskip}}]{${\tt branch}(\bullet)$}
  \begin{pgfscope}
    \pgfpathmoveto{\pgfpointxy{0.825}{-0.375}}
    \pgfpathlineto{\pgfpointxy{0.725}{-0.350}}
    \pgfpathlineto{\pgfpointxy{0.825}{-0.325}}
    \pgfpathlineto{\pgfpointxy{0.825}{-0.375}}
    \pgfsetfillcolor{black}
    \pgfsetlinewidth{0.100pt}
    \pgfusepath{fill,stroke}
  \end{pgfscope}
  \begin{pgfscope}
    \pgfpathmoveto{\pgfpointxy{0.475}{0.100}}
    \pgfpathlineto{\pgfpointxy{0.975}{0.100}}
    \pgfpathlineto{\pgfpointxy{0.975}{-0.350}}
    \pgfpathlineto{\pgfpointxy{0.825}{-0.350}}
    \pgfusepath{stroke}
  \end{pgfscope}
  \begin{pgfscope}
    \pgfpathmoveto{\pgfpointxy{0.825}{-0.555}}
    \pgfpathlineto{\pgfpointxy{0.725}{-0.530}}
    \pgfpathlineto{\pgfpointxy{0.825}{-0.505}}
    \pgfpathlineto{\pgfpointxy{0.825}{-0.555}}
    \pgfsetfillcolor{black}
    \pgfsetlinewidth{0.100pt}
    \pgfusepath{fill,stroke}
  \end{pgfscope}
  \begin{pgfscope}
    \pgfpathmoveto{\pgfpointxy{0.575}{-0.250}}
    \pgfpathlineto{\pgfpointxy{1.175}{-0.250}}
    \pgfpathlineto{\pgfpointxy{1.175}{-0.530}}
    \pgfpathlineto{\pgfpointxy{0.825}{-0.530}}
    \pgfusepath{stroke}
  \end{pgfscope}
  \begin{pgfscope}
    \pgfpathmoveto{\pgfpointxy{2.475}{-0.525}}
    \pgfpathlineto{\pgfpointxy{2.375}{-0.500}}
    \pgfpathlineto{\pgfpointxy{2.475}{-0.475}}
    \pgfpathlineto{\pgfpointxy{2.475}{-0.525}}
    \pgfsetfillcolor{black}
    \pgfsetlinewidth{0.100pt}
    \pgfusepath{fill,stroke}
  \end{pgfscope}
  \begin{pgfscope}
    \pgfpathmoveto{\pgfpointxy{2.475}{0.000}}
    \pgfpathlineto{\pgfpointxy{2.875}{0.000}}
    \pgfpathlineto{\pgfpointxy{2.875}{-0.500}}
    \pgfpathlineto{\pgfpointxy{2.475}{-0.500}}
    \pgfusepath{stroke}
  \end{pgfscope}
  \begin{pgfscope}
    \pgfpathmoveto{\pgfpointxy{2.775}{0.395}}
    \pgfpathlineto{\pgfpointxy{2.675}{0.420}}
    \pgfpathlineto{\pgfpointxy{2.775}{0.445}}
    \pgfpathlineto{\pgfpointxy{2.775}{0.395}}
    \pgfsetfillcolor{black}
    \pgfsetlinewidth{0.100pt}
    \pgfusepath{fill,stroke}
  \end{pgfscope}
  \begin{pgfscope}
    \pgfpathmoveto{\pgfpointxy{2.575}{-0.330}}
    \pgfpathlineto{\pgfpointxy{3.075}{-0.330}}
    \pgfpathlineto{\pgfpointxy{3.075}{0.420}}
    \pgfpathlineto{\pgfpointxy{2.775}{0.420}}
    \pgfusepath{stroke}
  \end{pgfscope}
\end{pgfpicture}
\ifx\Setlineno\undefined\else\Setlineno=32\fi
\end{center}

\caption{Shape of generated code for ${\tt if\ }e_1 = e_2{\tt \ then\ }c_1{\tt \ else\ }c_2$
(left) and \\ ${\tt while\ }e_1 = e_2{\tt \ do\ }c{\tt \ done}$ (right)}
\label{f:compilation}

\end{figure}

The code $\comp(e)$ for an expression evaluates $e$ and pushes its
value on top of the stack \coq{compile\_expr}.  It executes linearly
(no branches) and leaves the store unchanged.  (This is the familiar
translation from algebraic notation to reverse Polish notation.)
\begin{eqnarray*}
\comp(x) & = & {\tt var}(x) \\
\comp(n) & = & {\tt const}(n) \\
\comp(e_1 + e_2) & = & \comp(e_1);\comp(e_2);{\tt add} \\
\comp(e_1 - e_2) & = & \comp(e_1);\comp(e_2);{\tt sub}
\end{eqnarray*}
The code $\comp(b,\delta)$ for a boolean expression falls
through if $b$ is true, and branches to offset $\delta$ if $b$ is false.
\coq{compile\_bool\_expr}
\begin{eqnarray*}
\comp(e_1 = e_2,~\delta) & = & \comp(e_1);\comp(e_2);{\tt bne}(\delta) \\
\comp(e_1 < e_2,~\delta) & = & \comp(e_1);\comp(e_2);{\tt bge}(\delta)
\end{eqnarray*}
The code $\comp(c)$ for a command $c$ updates the state
according to the semantics of $c$, while leaving the stack unchanged.
\coq{compile\_cmd}
\begin{eqnarray*}
\comp({\tt skip}) & = & \epsilon \\
\comp(x := e) & = & \comp(e);{\tt setvar}(x) \\
\comp(c_1; c_2) & = & \comp(c_1);\comp(c_2) \\
\comp({\tt if\ }b{\tt \ then\ }c_1{\tt \ else\ }c_2) & = &
  \comp(b, \|C_1\|+1);C_1;{\tt branch}(\|C_2\|);C_2 \\
& & \mbox{where $C_1 = \comp(c_1)$ and $C_2 = \comp(c_2)$} \\
\comp({\tt while\ }b{\tt \ do\ }c{\tt \ done}) & = &
  B;C;{\tt branch}(-(\|B\| + \|C\| + 1)) \\
& & \mbox{where $C = \comp(c)$ and $B = \comp(b, \|C\|+1)$}
\end{eqnarray*}
$\|C\|$ is the length of a list of instructions $C$.  The mysterious
offsets in branch instructions are depicted in figure~\ref{f:compilation}.

\medskip

Finally, the compilation of a program $c$ is 
${\tt compile}(c) = \comp(c);{\tt halt}$. \coq{compile\_program}


\medskip

Combining the compilation scheme with the semantics of the virtual
machine, we obtain a new way to execute a program $c$ in initial state
$s$: start the machine in code $\comp(c)$ and state $(0, \epsilon, s)$
(program counter at first instruction of $\comp(c)$; empty stack;
state $s$), and observe its behavior.  Does this behavior agree with the
behavior of $c$ predicted by the semantics of section~\ref{s:semantics}?

\subsection{Notions of semantic preservation}

Consider two programs $P_1$ and $P_2$, possibly in different
languages. (For example, $P_1$ is an IMP command and $P_2$ a
sequence of VM instructions.)  Under which conditions can we say that
$P_2$ preserves the semantics of $P_1$?  

To make this question precise, we assume given operational semantics
for the two languages that associate to $P_1, P_2$
sets $\behaviors(P_1), \behaviors(P_2)$ of observable
behaviors.  In our case, observable behaviors are: termination on a
final state $s$, divergence, and ``going wrong''.  The set
$\behaviors(P)$ contains exactly one element if $P$ has deterministic
semantics, two or more otherwise.

Here are several possible formal characterizations of the informal
claim that $P_2$ preserves the semantics of $P_1$.
\begin{itemize}
\item Bisimulation (equivalence):
         $\behaviors(P_1) = \behaviors(P_2)$
\item Backward simulation (refinement):
         $\behaviors(P_1) \supseteq \behaviors(P_2)$
\item Backward simulation for correct source programs:
         if ${\tt wrong} \notin \behaviors(P_1)$ then
         $\behaviors(P_1) \supseteq \behaviors(P_2)$
\item Forward simulation:
         $\behaviors(P_1) \subseteq \behaviors(P_2)$
\item Forward simulation for correct source programs:
         if ${\tt wrong} \notin \behaviors(P_1)$ then
         $\behaviors(P_1) \subseteq \behaviors(P_2)$
\end{itemize}

Bisimulation is the strongest notion of semantic preservation,
ensuring that the two programs are indistinguishable.  It is often too
strong in practice.  For example, the C language has non-deterministic
semantics because the evaluation order for expressions is not fully
specified; yet, C compilers choose one particular evaluation order
while generating deterministic machine code; therefore, the generated
code has fewer behaviors than the source code.  This intuition
corresponds to the backward simulation property defined above: all
behaviors of $P_2$ are possible behaviors of $P_1$, but $P_1$ can have
more behaviors.

In addition to reducing nondeterminism, compilers routinely optimize
away ``going wrong'' behaviors.  For instance, the source program
$P_1$ contains an integer division $z := x / y$ that can go wrong if
$y = 0$, but the compiler eliminated this division because $z$ is not
used afterwards, therefore generating a program $P_2$ that does not go
wrong if $y = 0$.  This additional degree of liberty is reflected in
the ``backward simulation for correct source programs'' above.

Finally, the two ``forward simulation'' properties reverse the roles
of $P_1$ and $P_2$, expressing the fact that any (non-wrong) behavior
of the source program $P_1$ is a possible behavior of the compiled
code $P_2$.  Such forward simulation properties are generally much easier to
prove than backward simulations, but provide apparently weaker
guarantees: $P_2$ could have additional behaviors, not exhibited by
$P_1$, that are undesirable, such as ``going wrong''.  This cannot
happen, however, if $P_2$ has deterministic semantics.

\begin{lemma} (Simulation and determinism.)
If $P_2$ has deterministic semantics,
then ``forward simulation for correct programs'' implies
``backward simulation for correct programs''.
\end{lemma}

In conclusion, for deterministic languages such as IMP and IMP virtual
machine code, ``forward simulation for correct programs'' is an
appropriate notion of semantic preservation to prove the correctness
of compilers and program transformations. 

\subsection{Semantic preservation for the compiler}

Recall the informal specification for the code $\comp(e)$ generated by
the compilation of expression $e$: it should evaluate $e$ and push its
value on top of the stack, execute linearly (no branches), and leave
the store unchanged.  Formally, we should have
 $\comp(e):~ (0, \sigma, s) \redm (\|\comp(e)\|, (\den{e}~s).\sigma, s)$
for all stacks~$\sigma$ and stores~$s$.  Note that $pc = \|\comp(e)\|$
means that the program counter is one past the last instruction in the
sequence $\comp(e)$.  To enable a proof by induction, we need to
strengthen this result and consider codes of the form $C_1; \comp(e); C_2$,
where the code for $e$ is bracketed by two arbitrary code sequences
$C_1$ and $C_2$.  The program counter, then, should go from
$\|C_1\|$ (pointing to the first instruction of $\comp(e)$) to
$\|C_1\| + \|\comp(e)\|$ (pointing one past the last instruction of
$\comp(e)$, or equivalently to the first instruction of $C_2$).

\begin{lemma} \coq{compile\_expr\_correct} \label{compile-expr-correct}
For all instruction sequences $C_1, C_2$, stacks~$\sigma$ and states~$s$,
$$ C_1;\comp(e);C_2 |-
   (\|C_1\|, \sigma, s) \redm (\|C_1\| + \|\comp(e)\|, \den{e}~s.\sigma, s) $$
\end{lemma}

The proof is a simple induction on the structure of $e$.  Here is a
representative case: $e = e_1 + e_2$.  Write $v_1 = \den{e_1}~s$ and
$v_2 = \den{e_2}~s$.  The code $C$ is $C_1; \comp(e_1); \comp(e_2);
{\tt add}; C_2$.  Viewing $C$ as $C_1; \comp(e_1);
(\comp(e_2);{\tt add};C_2)$, we can apply the induction hypothesis to
$e_1$, obtaining the transitions
$$ (\|C_1\|, \sigma, s) \redm (\|C_1\| + \|\comp(e_1)\|, v_1.\sigma, s) $$
Likewise, viewing $C$ as
$(C_1; \comp(e_1)); \comp(e_2); ({\tt add};C_2)$, we can apply the
induction hypothesis to $e_2$, obtaining
$$ (\|C_1; \comp(e_1)\|, v_1.\sigma, s) \redm
  (\|C_1; \comp(e_1)\| + \|\comp(e_2)\|, v_2.v_1.\sigma, s) $$
Combining these two sequences with an {\tt add} transition, we obtain
$$ (\|C_1\|, \sigma, s) \redm
  (\|C_1; \comp(e_1);\comp(e_2)\| + 1, (v_1+v_2).\sigma, s) $$
which is the desired result.


The statement and proof of correctness for the compilation of
boolean expressions is similar.  Here, the stack and the
store are left unchanged, and control is transferred either to the end
of the generated instruction sequence or to the given offset relative
to this end, depending on the truth value of the condition.

\begin{lemma} \coq{compile\_bool\_expr\_correct}
For all instruction sequences $C_1, C_2$, stacks~$\sigma$ and states~$s$,
$$ C_1;\comp(b,~\delta);C_2 |-
   (\|C_1\|, \sigma, s) \redm (pc, \sigma, s)$$
with $pc = \|C_1\| + \|\comp(b)\|$ if $\den{b}~s = {\tt true}$
and $pc = \|C_1\| + \|\comp(b)\| + \delta$ otherwise.
\end{lemma}

To show semantic preservation between an IMP command and its compiled
code, we prove a ``forward simulation for correct programs'' result.
We therefore have two cases to consider: (1) the command terminates
normally, and (2) the command diverges.  In both cases, we use the
natural semantics to conduct the proof, since its compositional nature
is a good match for the compositional nature of the compilation scheme.  

\begin{theorem} \coq{compile\_cmd\_correct\_terminating}
Assume $\eval s c {s'}$.  Then, for all instruction sequences $C_1, C_2$
and stack $\sigma$,
$$ C_1;\comp(c);C_2 |-
   (\|C_1\|, \sigma, s) \redm (\|C_1\| + \|\comp(c)\|, \sigma, s') $$
\end{theorem}

The proof is by induction on a derivation of $\eval s c {s'}$ and
uses the same techniques as that of lemma~\ref{compile-expr-correct}.

For the diverging case, we need the following special-purpose
coinduction principle.

\begin{lemma}
Let $X$ be a set of (machine code, machine state) pairs such that
$$ \forall (C,S) \in X, ~ \exists S', ~(C, S') \in X \wedge C |- S \redp S'. $$
Then, for all $(C, S) \in X$, we have $C |- S \diverges$
(there exists an infinite sequence of transitions starting from $S$).
\end{lemma}

\noindent
The following theorem follows from the coinduction principle above
applied to the set
$$X = \{ (C_1;\comp(c);C_2, (\|C_1\|, \sigma, s)) \mid \evalinf s c \}.$$

\begin{theorem} \coq{compile\_cmd\_correct\_diverging}
Assume $\evalinf s c$.  Then, for all instruction sequences 
$C_1, C_2$ and stacks $\sigma$,
$$ C_1;\comp(c);C_2 |-
   (\|C_1\|, \sigma, s) \diverges
$$
\end{theorem}

This completes the proof of forward simulation for correct programs.

\subsection{Further reading}

The virtual machine used in this section matches a small subset of the
Java Virtual Machine \cite{JVMspec}.  Other examples of mechanized
verification of nonoptimizing compilers producing virtual machine code
include Bertot \cite{Bertot-98} (for the IMP language),
Klein and Nipkow \cite{Klein-Nipkow-jinja} (for a subset of Java),
and Grall and Leroy \cite{Leroy-Grall-coindsem} (for call-by-value
$\lambda$-calculus).  The latter two show forward simulation results;
Bertot shows both forward and backward simulation, and concludes that
backward simulation is considerably more difficult to prove.  Other
examples of difficult backward simulation arguments (not mechanized) can
be found in \cite{Hardin-Maranget-Pagano}, for call-by-name and
call-by-value $\lambda$-calculus.

Lemma~\ref{compile-expr-correct} (correctness of compilation of
arithmetic expression to stack machine code) is historically
important: it is the oldest published compiler correctness proof
(McCarthy and Painter \cite{McCarthy-Painter-67}, in 1967) and the
oldest mechanized compiler correctness proof (Milner and Weyhrauch,
\cite{Milner-compiler-correctness}, in 1972).  Since then, a great
many correctness proofs for compilers and compilation passes have been
published, some of them being mechanized: Dave's bibliography
\cite{Dave-compiler-verif-03} lists 99 references up to 2002.

\section{An example of optimizing program transformation: dead code elimination}

Compilers are typically structured as a sequence of program
transformations, also called {\em passes}.  Some passes translate from
one language to another, lower-level language, closer to machine
code.  The compilation scheme of section~\ref{s:compilation} is a
representative example.  Other passes are optimizations: they rewrite
the program to an equivalent, but more efficient program.  For
example, the optimized program runs faster, or is smaller.  

In this section, we study a representative optimization: dead code
elimination.  The purpose of this optimization, performed on the IMP
source language, is to remove assignments $x := e$
(turning them into {\tt skip} instructions) such that the value of $x$ is
not used in the remainder of the program.  This reduces both the
execution time and the code size.

\begin{example} Consider the command {\tt x\ :=\ 1;\ y\ :=\ y\ +\ 1;\ x\ :=\ 2}.
The assignment {\tt x\ :=\ 1} can always be eliminated since {\tt x} is not
referenced before being redefined by {\tt x\ :=\ 2}.  
\end{example}

To detect the fact that the value of a variable is not used later, we
need a static analysis known as {\em liveness analysis}.  

\subsection{Liveness analysis}

A variable is {\em dead} at a program point if its value is not used
later in the execution of the program: either the variable is never
mentioned again, or it is always redefined before further use.  A
variable is {\em live} if it is not dead.

Given a set $A$ of variables live ``after'' a command $c$,
the function ${\tt live}(c,A)$ over-approximates the set of variables live
``before'' the command \coq{live}.
It proceeds by a form of reverse execution of
$c$, conservatively assuming that conditional branches can go
both ways. $FV$ computes the set of variables referenced in an expression
\coq{fv\_expr} \coq{fv\_bool\_expr}.

\begin{eqnarray*}
{\tt live}({\tt skip}, A) & = & A \\
{\tt live}(x := e, ~A) & = & 
  \cases{ (A \setminus \{x\}) \union FV(e) & if $x \in A$; \cr
          A & if $x \notin A$. \cr} \\
{\tt live}((c_1;c_2),~A) & = & {\tt live}(c_1, {\tt live}(c_2, A)) \\
{\tt live}(({\tt if\ }b{\tt \ then\ }c_1{\tt \ else\ }c_2), ~A) & = &
FV(b) \union {\tt live}(c_1, A) \union {\tt live}(c_2, A) \\
{\tt live}(({\tt while\ }b{\tt \ do\ }c{\tt \ done}), ~A) & = &
{\tt fix}(\lambda X.~A \union FV(b) \union {\tt live}(c, X))
\end{eqnarray*}

If $F$ is a function from sets of variables to sets of variables,
${\tt fix}(F)$ is supposed to compute a {\em post-fixpoint} of $F$,
that is, a set $X$ such that $F(X) \subseteq X$.  Typically, $F$ is
iterated $n$ times, starting from the empty set, until we reach an $n$
such that $F^{n+1}(\emptyset) \subseteq F^n(\emptyset)$.  
Ensuring termination of such an iteration is, in general, a difficult
problem.  (See section~\ref{s:optim-further} for discussion.)
To keep things simple, we bound arbitrarily to $N$ the
number of iterations, and return a default over-approximation if a
post-fixpoint cannot be found within $N$ iterations: \coq{fixpoint}
$$
{\tt fix}(F, {\it default}) = 
\cases{F^n(\emptyset) &
      if $\exists n \le N, ~ F^{n+1}(\emptyset) \subseteq F^n(\emptyset)$;
      \cr
      {\it default} & otherwise \cr}
$$
Here, a suitable default is 
$A \union FV({\tt while\ }b{\tt \ do\ }c{\tt \ done})$, the set of variables
live ``after'' the loop or referenced within the loop.
\begin{eqnarray*}
{\tt live}(({\tt while\ }b{\tt \ do\ }c{\tt \ done}), ~A) & = &
{\tt fix}(\begin{array}[t]{@{}l}
      \lambda X.~A \union FV(b) \union {\tt live}(c, X), \\
      A \union FV({\tt while\ }b{\tt \ do\ }c{\tt \ done}))
      \end{array}
\end{eqnarray*}

\begin{lemma} \coq{live\_while\_charact}
Let $A' = {\tt live}({\tt while\ }b{\tt \ do\ }c{\tt \ done},~A)$.  Then:
$$ FV(b) \subseteq A' \qquad A \subseteq A' \qquad 
   {\tt live}(c, A') \subseteq A' $$
\end{lemma}

\subsection{Dead code elimination}

The program transformation that eliminates dead code is, then: \coq{dce}
\begin{eqnarray*}
{\tt dce}({\tt skip}, A) & = & {\tt skip} \\
{\tt dce}(x := e, ~A) & = & 
  \cases{ x := e & if $x \in A$; \cr
          {\tt skip} & if $x \notin A$. \cr} \\
{\tt dce}((c_1; c_2),~A) & = &
  {\tt dce}(c_1, {\tt live}(c_2, A)); {\tt dce}(c_2, A) \\
{\tt dce}(({\tt if\ }b{\tt \ then\ }c_1{\tt \ else\ }c_2), ~A) & = &
{\tt if\ }b{\tt \ then\ }{\tt dce}(c_1, A){\tt \ else\ }{\tt dce}(c_2, A) \\
{\tt dce}({\tt while\ }b{\tt \ do\ }c{\tt \ done}, ~A) & = &
{\tt while\ }b{\tt \ do\ }{\tt dce}(c, A){\tt \ done}
\end{eqnarray*}

\begin{example} Consider again the ``Euclidean division'' program $c$:
\begin{verbatim}
    r := a; q := 0;  while b < r+1 do r := r - b; q := q + 1 done
\end{verbatim}
If {\tt q} is not live ``after'' (${\tt q} \notin A$), it is not live
throughout this program either.  Therefore, ${\tt dce}(c,A)$ produces
\begin{verbatim}
    r := a; skip;    while b < r+1 do r := r - b; skip       done
\end{verbatim}
The useless computations of {\tt q} have been eliminated entirely, in a
process similar to program slicing.  In contrast, if {\tt q} is live
``after'' (${\tt q} \in A$), all computations are necessary and
${\tt dce}(c,A)$ returns $c$ unchanged.

\end{example}

\subsection{Correctness of the transformation}

We show a ``forward simulation for correct programs'' property:
\begin{itemize}
\item If $c, s \terminates s'$, then ${\tt dce}(c,A), s \terminates s''$
for some $s''$ related to $s'$.
\item If $c, s \diverges$, then ${\tt dce}(c,A), s \diverges$.
\end{itemize}
However, the program ${\tt dce}(c,A)$ performs fewer assignments than
$c$, therefore the final states can differ on the values of dead
variables.  We define agreement between two states $s, s'$ with
respect to a set of live variables $A$. \coq{agree}
$$ \agree{s}{s'}{A} \defequal \forall x \in A,~s(x) = s'(x) $$

\begin{lemma} \coq{eval\_expr\_agree} \coq{eval\_bool\_expr\_agree}
Assume $\agree{s}{s'}{A}$.
If $FV(e) \subseteq A$, then $\den{e}~s = \den{e}~s'$.
If $FV(b) \subseteq A$, then $\den{b}~s = \den{b}~s'$.
\end{lemma}

The following two key lemmas show that agreement is preserved by
parallel assignment to a live variable, or by unilateral assignment to
a dead variable.  The latter case corresponds to the replacement of $x
:= e$ by {\tt skip}.

\begin{lemma} \coq{agree\_update\_live} (Assignment to a live variable.)
If $\agree{s}{s'}{A \setminus \{x\}}$, then
$\agree{s[x \becomes v]}{s'[x \becomes v]}{A}$.
\end{lemma}

\begin{lemma} \coq{agree\_update\_dead} (Assignment to a dead variable.)
If $\agree{s}{s'}{A}$ and $x \notin A$, then
$\agree{s[x \becomes v]}{s'}{A}$.
\end{lemma}

Using these lemmas, we can show forward simulation diagrams both for
terminating and diverging commands $c$.  In both case, we assume agreement
on the variables live ``before'' $c$, namely ${\tt live}(c, A)$.

\begin{theorem} \coq{dce\_correct\_terminating}
If $\eval s c {s'}$ and $\agree s {s_1} {{\tt live}(c, A)}$, then
there exists $s_1'$ such that $\eval {s_1} {{\tt dce}(c,A)} {s'_1}$
and $\agree {s'} {s_1'} {A}$.
\end{theorem}

\begin{theorem} \coq{dce\_correct\_diverging}
If $\evalinf s c$ and $\agree s {s_1} {{\tt live}(c, A)}$, then
$\evalinf {s_1} {{\tt dce}(c,A)}$.
\end{theorem}

\subsection{Further reading} \label{s:optim-further}

Dozens of compiler optimizations are known, each targeting a
particular class of inefficiencies.  See Appel \cite{Appel-tiger} for
an introduction to optimization, and Muchnick
\cite{Muchnick} for a catalog of classic optimizations.

The results of liveness analysis can be exploited to perform register
allocation (a crucial optimization performance-wise), following
Chaitin's approach \cite{Chaitin-82} \cite[chap.~11]{Appel-tiger}:
coloring of an interference graph.  A mechanized proof of correctness
for graph coloring-based register allocation,
extending the proof given in this section, is described by Leroy
\cite{Leroy-Compcert-CACM,Leroy-backend-08}.

Liveness analysis is an instance of a more general class of static
analyses called {\em dataflow analyses} \cite[chap.~17]{Appel-tiger},
themselves being a special case of abstract interpretation.  Bertot
\etal \cite{Bertot-Gregoire-Leroy-05} and Leroy
\cite{Leroy-backend-08} prove, in Coq, the correctness of several
optimizations based on dataflow analyses, such as constant propagation and
common subexpression elimination.  Cachera \etal \cite{Cachera-Jensen-05}
present a reusable Coq framework for dataflow analyses.

Dataflow analyses are generally carried on an unstructured
representation of the program called the control-flow graph.  Dataflow
equations are set up between the nodes of this graph, then solved by
one global fixpoint iteration, often based on Kildall's worklist
algorithm \cite{Kildall-73}.  This is more efficient than the approach
we described (computing a local fixpoint for each loop), which can be
exponential in the nesting degree of loops.  Kildall's worklist
algorithm has been mechanically verified many times
\cite{Bertot-Gregoire-Leroy-05,Coupet-Grimal-Delobel-05,Klein-Nipkow-jinja}.

The effective computation of fixpoints is a central issue in static
analysis.  Theorems such as Knaster-Tarski's show the existence of
fixpoints in many cases, and can be mechanized
\cite{Paulson-95,Bertot-Komendantsky}, but fail to provide effective
algorithms.  Noetherian recursion can be used if the domain of the analysis
is well founded (no infinite chains)
\cite[chap.~15]{Bertot-Casteran-Coqart}, but this property is
difficult to ensure in practice \cite{Cachera-Jensen-05}.  The
shortcut we took in this section (bounding arbitrarily the number of
iterations) is inelegant but a reasonable engineering compromise.

\section{State of the art and current trends}

While this lecture was illustrated using ``toy'' languages and machines,
the techniques we presented, based on operational and axiomatic
semantics and on their mechanization using proof assistants, do scale
to realistic programming languages and systems.  Here are some recent
achievements using similar techniques, in reverse chronological order.

\begin{itemize}
\item The verification of the seL4 secure micro-kernel
(\url{http://nicta.com.au/research/projects/l4.verified/})
\cite{Klein-09}.

\item The CompCert verified compiler: a realistic,
  moderately-optimizing compiler for a large subset of the C language
  down to PowerPC and ARM assembly code.
(\url{http://compcert.inria.fr/})
\cite{Leroy-Compcert-CACM}.

\item The Verisoft project 
(\url{http://www.verisoft.de/}), which aims
at the end-to-end formal verification of a complete embedded system,
from hardware to application.

\item Formal specifications of the Java / Java Card virtual machines
  and mechanized verifications of the Java bytecode verifier:
  Ninja \cite{Klein-Nipkow-jinja},
  Jakarta \cite{Barthe-02},
  Bicolano (\url{http://mobius.inria.fr/twiki/bin/view/Bicolano}),
  and the Kestrel Institute project
  (\url{http://www.kestrel.edu/home/projects/java/}).

\item Formal verification of the ARM6 processor micro-architecture
  against the ARM instruction set specification \cite{Fox-ARM6-03}

\item The ``foundational'' approach to Proof-Carrying Code
  \cite{Appel-FPCC}.

\item The CLI stack: a formally verified microprocessor and
  compiler from an assembly-level language
(\url{http://www.cs.utexas.edu/~moore/best-ideas/piton/index.html})
\cite{Moore-96}.

\end{itemize}

\noindent Here are some active research topics in this area.

\paragraph{Combining static analysis and program proof.}  Static analysis
  can be viewed as the automatic generation of logical assertions,
  enabling the results of static analysis to be verified {\em a
  posteriori} using a program logic, and facilitating the annotation
  of existing code with logical assertions.

\paragraph{Proof-preserving compilation.}  Given a source program annotated
  with assertions and a proof in axiomatic semantics, can we produce
  machine code annotated with the corresponding assertions and the
  corresponding proof?  \cite{Barthe-Gregoire-06,Li-Owens-Slind-07}.

\paragraph{Binders and $\alpha$-conversion.}
  A major obstacle to the mechanization of rich language semantics
  and advanced type systems is the handling of bound variables and the
  fact that terms containing binders are equal modulo
  $\alpha$-conversion of bound variables.  The POPLmark challenge
  explores this issue \cite{POPLmark}.

\paragraph{Shared-memory concurrency.}
  Shared-memory concurrency raises major
  semantic difficulties, ranging from formalizing the
  ``weakly-consistent'' memory models implemented by today's multicore
  processors \cite{Sewell-POPL09} to mechanizing program logics
  appropriate for proving concurrent programs correct
  \cite{Feng-Ferreira-Shao,Hobor-Appel-Zappa-08}.

\paragraph{Progressing towards fully-verified development and verification
  environments for high-assurance software.}  Beyond verifying
  compilers and other code generation tools, we'd like to gain formal
  assurance in the correctness of program verification tools such as
  static analyzers and program provers.


\bibliographystyle{abbrv}
\ifx\french\undefined \def\biling#1#2{#1} \else \def\biling#1#2{#2}
  \fi\ifx\abbrevbib\undefined \def\abbrev#1#2{#1} \else \def\abbrev#1#2{#2} \fi

\end{document}